\documentclass[leqno, 12pt]{article}
\usepackage{bbm}
\usepackage{fnpct}
\usepackage[multiple, bottom]{footmisc}
\usepackage{mathtools}

\def\clap#1{\hbox to 0pt{\hss#1\hss}}

\usepackage{booktabs,caption}
\usepackage[flushleft]{threeparttable}
\usepackage{bm}
\usepackage{dcolumn}
\usepackage{amsfonts}
\usepackage{amsmath}
\usepackage{amssymb}
\usepackage{nccmath}
\usepackage[english]{babel}
\usepackage{accents}
\usepackage{natbib}
\usepackage{url}
\usepackage{array}    
\usepackage{alphalph}
\usepackage{setspace} 
\usepackage{graphicx}
\usepackage[top=25truemm, bottom=25truemm, left=25truemm, right=25truemm]{geometry}

\usepackage{float} 

\usepackage{etoolbox}
\patchcmd{\MaketitleBox}{\footnotesize\itshape\elsaddress\par\vskip36pt}{\footnotesize\itshape\elsaddress\par\parbox[b][36pt]{\linewidth}{\vfill\hfill\textnormal{This version: \today}\hfill\null\vfill}}{}{}%
\patchcmd{\pprintMaketitle}{\footnotesize\itshape\elsaddress\par\vskip36pt}{\footnotesize\itshape\elsaddress\par\parbox[b][36pt]{\linewidth}{\vfill\hfill\textnormal{\today}\hfill\null\vfill}}{}{}%

\begin{document}

\thispagestyle{empty}
\renewcommand{\thefootnote}{\fnsymbol{footnote}}

\begin{center}
{\LARGE Economic Integration and Agglomeration of Multinational Production with Transfer Pricing\footnote{We wish to thank Co-Editor, Masaki Aoyagi, and three anonymous referees for valuable comments and suggestions.
This paper is a much revised version of Chapter 2 of Okoshi's Ph.D. thesis submitted to the University of Munich (\citealp{OkoshiPHD}).
Helpful comments from the committee members, Carsten Eckel, Andreas Haufler and Dominika Langenmayr are greatly appreciated.
We are grateful to Cristina Angelico, Makoto Hasegawa, Tsung-Yu Ho, Yukio Karasawa-Ohtashiro and Dirk Schindler for extensive discussions.
Thanks also to participants of seminars and conferences at Kagawa U, Kochi U, U of Munich, IIPF (U of Tampere), 2019 Symposium of Public Economics (Osaka U), APTS  (U of Tokyo), Osaka U (ISER and OSIPP), U of Tokyo, ERSA (U of Lyon), MYEM Conference (U of Munich), Chulalongkorn U, Kobe U, Hanyang U, the Third International Workshop ``Market Studies and Spatial Economics'' (HSE), and NASMES (UQAM) for comments.  
Financial supports from the Japan Society for the Promotion of Science (Grant Numbers: JP16J01228;  JP17K03789; JP18H00866; JP19K13693; JP99K13693; JP20K22122), the German Research
Foundation through GRK1928, the Nomura Foundation, and the Obayashi Foundation are gratefully acknowledged.}}

\

\

{\large Hayato Kato\footnote{Corresponding author. Graduate School of Economics, Osaka University, 1-7 Machikaneyama, Toyonaka, Osaka, 560-0043, Japan.
{\it E-mail address:} \ hayato.kato@econ.osaka-u.ac.jp} 
\ \ \ \ Hirofumi Okoshi\footnote{Faculty of Economics, Okayama University, 3-1-1 Tsushima, Kita-ku, Okayama, 700-8530, Japan.
{\it E-mail address:} \ hirofumi.okoshi1@gmail.com} }
\end{center}

\begin{center}
\today
\end{center}

\

\vspace{0.25cm}

\begin{abstract}
Do low corporate taxes always favor multinational production over economic integration?
We propose a two-country model in which multinationals choose the locations of production plants and foreign distribution affiliates and shift profits between them through transfer prices.
With high trade costs, plants are concentrated in the low-tax country; surprisingly, this pattern reverses with low trade costs.
Indeed, economic integration has a non-monotonic impact: falling trade costs first decreases and then increases the plant share in the high-tax country, which we empirically confirm.
Moreover, allowing for transfer pricing makes tax competition tougher and international coordination on transfer-pricing regulation can be beneficial.
\end{abstract}



\renewcommand{\thefootnote}{\arabic{footnote}}
\setcounter{footnote}{0}
\setcounter{page}{0}

\pagebreak

\renewcommand{\thefootnote}{\arabic{footnote}}
\setcounter{footnote}{0}

\

\section{Introduction}

Progressive economic integration in the last few decades has increased the international mobility of multinational enterprises (MNEs), allowing them to diversify activities across subsidiaries in different countries.
Considering the complexity of multinational activities,
governments today need to carefully design policies to attract MNEs.
Among many factors, corporate taxation is one of the essential determinants of foreign direct investment (FDI) (\citealp[Chapter 6]{NavarettiVenables2004}; \citealp{BlonigenPiger2014}).\footnote{As for other determinants of MNEs' location decision, recent studies highlight agglomeration economies arising from affiliates (\citealp{Mayeretal2010}) and financial development in the host country (\citealp{Biliretal2019}).}
One naturally expects that countries with a low corporate tax rate would succeed in hosting more FDI inflows than those with a high tax rate.
Earlier empirical studies confirmed this using data on FDI in {\it all sectors} (e.g., \citealp{Benassyetal2005}; \citealp{Eggeretal2009}).

However, the type of activities of multinationals that operate in such low-tax countries is not obvious.
Governments reduce taxes to attract production plants, which contribute to local employment and tax revenues.\footnote{The Irish government, for example, has explicitly stated its commitment to the low corporate tax rate to attract FDI. See the 2013 Financial Statement by the Minister for Finance: \url{http://www.budget.gov.ie/Budgets/2013/FinancialStatement.aspx}, accessed on November 25, 2020.}
Contrary to host governments' expectations, MNEs may establish affiliates in low-tax countries just to save taxes and may not engage in production (\citealp{HornerAoyama2009}).\footnote{\cite{HornerAoyama2009} provide a list of Ireland-based MNEs' relocations.
There are several examples where some MNEs moved production from Ireland abroad while maintaining non-production activities, such as service centers and marketing, in Ireland.}
As economic integration has dismantled barriers to goods' and factors' mobility in recent years, MNEs may put more emphasis on other barriers such as high taxes when choosing a location.

Figure 1 illustrates this point and shows that countries with lower taxes do not necessarily attract more multinational production.
In Figure 1(a), we take 23 Organisation for Economic Co-operation and Development (OECD) countries and draw the relationship between each country's average corporate tax rate from 2008 to 2016 and the average number of foreign affiliates in all sectors coming from the other OECD countries in the same period.\footnote{The sample countries do not include four European countries identified as tax havens by \cite{Zucman2014}, i.e., Ireland, Luxembourg, the Netherlands, and Switzerland.
The patterns laid out in Figure 1 remain unchanged even after including these four countries.}
To control for the host country's size, the average number of affiliates is divided by the host country's average GDP during the sample period.
The fitted line with a clear downward slope tells us that countries with a lower tax rate tend to attract more MNEs.
Figure 1(b) shows the type of MNEs' activities by plotting the share of affiliates in manufacturing sectors out of those in all sectors.
The fitted line has little explanatory power, with $R$-squared being $0.088$, and its slope has little statistical significance.
This suggests that whether tax rates are high or low does not contribute much to the share of foreign manufacturing affiliates.

\

\begin{center}
\includegraphics[scale=0.5]{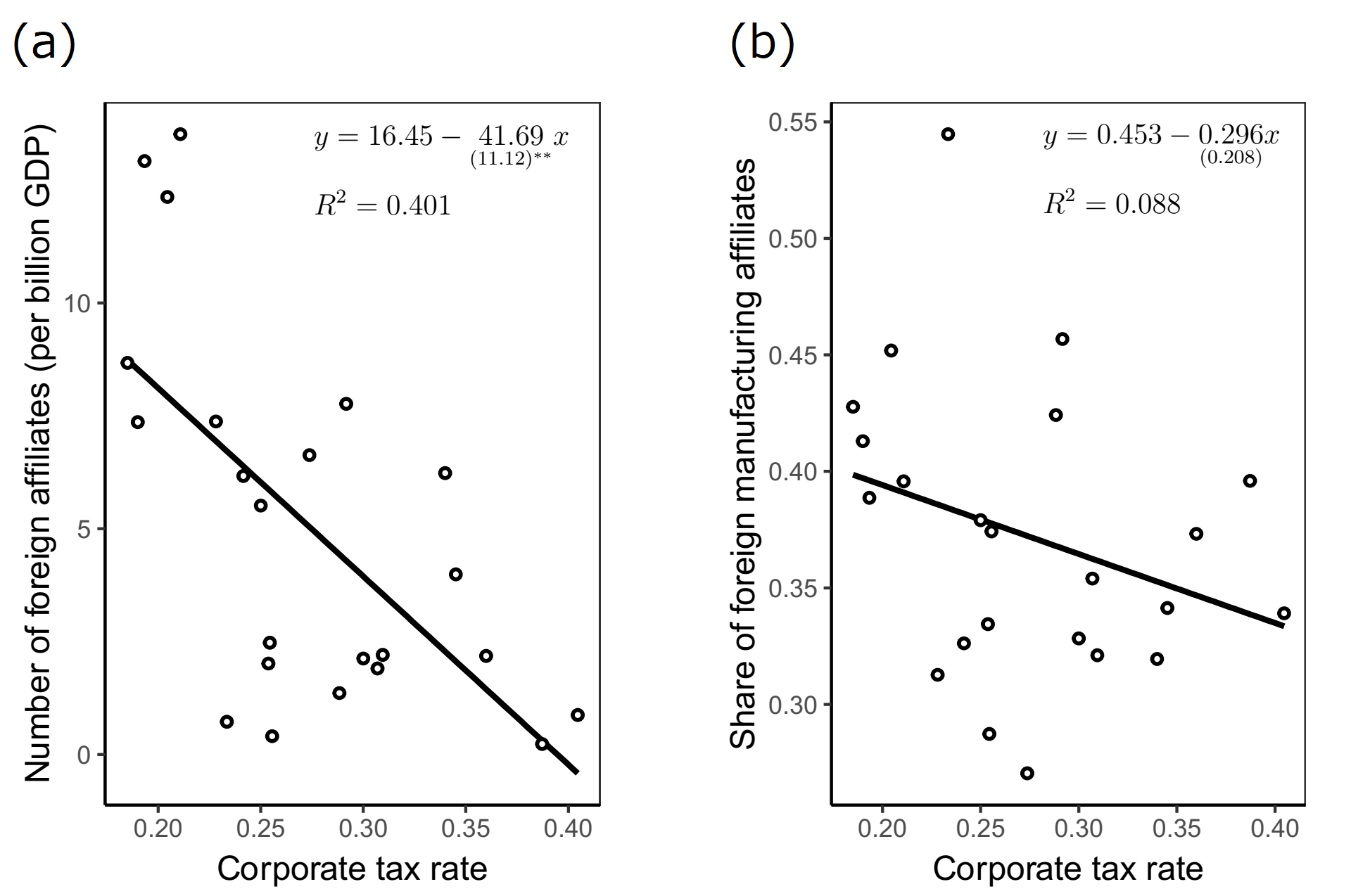} \\
Figure 1. \ Corporate tax rates and foreign affiliates in 2008 to 2016
\end{center}
\begin{spacing}{1}{\footnotesize \noindent
{\it Sources}: \ Centre for Business Taxation Tax Database 2017 (\citealp{Habu2017}); Outward activity of multinationals by country of location--ISIC Rev 4. in OECD Statistics: \url{https://stats.oecd.org/} \\
{\it Notes}: \ The horizontal axis is the average statutory corporate tax rate of a country in 2008 to 2016.
In panel (a), the vertical axis is the average number of foreign affiliates in all sectors (per USD 1 billion GDP of the host country) coming from the other sample countries in 2008 to 2016.
In panel (b), it is the average share of foreign manufacturing affiliates out of those in all sectors.
Foreign affiliates in 2008 to 2016.
Numbers in parentheses below the fitted equation are standard errors, where ** indicates significance at the 5$\%$ level.
See Appendix 2 for details on data.}
\end{spacing}

\

This lack of a clear relationship between taxes and the manufacturing affiliate share can be explained by profit shifting by MNEs.
MNEs allocate their activities between low- and high-tax countries, and transfer profits by controlling prices for intra-firm trade, known as {\it transfer prices}.\footnote{Many studies provide empirical evidence on transfer pricing. See \cite{Swenson2001}; \cite{BartelsmanBeestma2003}; \cite{Clausing2003}; \cite{Bernardetal2006}; \cite{CristeaNguyen2016}; \cite{Gumpertetal2016}; \cite{Guvenenetal2017}; \cite{Bruneretal2018}; and \cite{Daviesetal2018}.}
For example, headquarters in high-tax countries make profits by selling goods to affiliates in low-tax countries by setting low transfer prices to inflate the affiliates' profits.
This profit shifting through intra-firm trade has been made easier by the recent proliferation of trade liberalization and transportation technology advancements.

When profits are transferable between countries with different tax rates, it is unclear where MNEs optimally set up their plants and affiliates.
To answer the question, we extend a two-country spatial model developed by \cite{MartinRogers1995}; and \cite{Pfluger2004} to incorporate MNEs with profit-shifting motives.

Specifically, we investigate in which country---the low-tax or high-tax one---multinational production is agglomerated and how the location pattern changes as trade costs fall.
There is a fixed mass of monopolistically competitive MNEs in the world, with each locating a plant (or headquarters) for production in one country and an affiliate for distribution in the other.
The MNE engages in intra-firm trade by exporting the output produced in the home country to the affiliate in a foreign country.
It can use the transfer price of the output for profit shifting.
However, due to trade costs, shipping goods from one country to another will be costly.
Trade costs change the volume of intra-firm trade, and thus, affect the effectiveness of profit shifting, which in turn affects the choice of MNEs' location.

Based on this setting, we obtain the following results.
With a low level of economic integration marked by high trade costs,
the low-tax country attracts a higher share of multinational production than the high-tax country does.
When high trade costs hamper intra-firm trade, thereby limiting the profit-shifting opportunity, MNEs can sell little to their foreign affiliates. 
As most of the profits are made in the country where goods are produced, 
they simply prefer to locate production in the low-tax country.

However, with a high level of economic integration marked by low trade costs, this location pattern reverses: production plants agglomerate in the high-tax country.
This result seems surprising, but is indeed consistent with MNEs' optimal location choice.
The MNE with production in the high-tax country {\it lowers} the transfer price to shift its home plant's profits to its foreign affiliate in the low-tax country.
The lowered transfer price reduces the affiliate's marginal cost, which allows it to lower the price of goods and gain competitiveness against local plants.
Conversely, the MNE with production in the low-tax country {\it raises} the transfer price to shift profits from its foreign affiliate in the high-tax country back to its home plant.
Due to the high transfer price, the affiliate sells goods at a high price and loses competitiveness against local plants.
Thus, transfer pricing favors the MNE with production in the high-tax country such that it makes them competitive in both home and foreign markets.
When trade costs are so low that this effect is significant,
all MNEs strategically choose to locate production in the high-tax country.

Moreover, the location pattern of multinational production is indeed non-monotonic.
That is, a fall in trade costs first decreases and then increases the share of production plants in the high-tax country.
A simple intuition that production is agglomerated in the low-tax country to save both taxes and trade costs holds under high trade costs; however, it does not apply under low trade costs, as transfer pricing  affects the strategic location choice of MNEs that seek price competitiveness.

These results may explain why taxes do not have a strong explanatory power for the share of manufacturing affiliates, as Figure 1 suggests.
In addition, \cite{Overesch2009} provides  empirical evidence that MNEs in high-tax Germany increase real investments as the cross-country corporate tax difference between their home country and Germany is larger.
Our own empirical exercise using the same data as those in Figure 1 also confirms the non-monotonic impact of economic integration on the distribution of multinational production.

As a result of transfer pricing, the high-tax country attracts more multinational production but does not enjoy greater tax revenues than it would without transfer pricing.
The opposite is true for the low-tax country.
In fact, allowing for transfer pricing lowers global tax revenues.
Amid growing concerns about tax base erosion,
the OECD recently reported that the estimated revenue losses from MNEs' tax avoidance are about 10$\%$ of global corporate income tax revenues.\footnote{
See \url{http://www.oecd.org/ctp/oecd-presents-outputs-of-oecd-g20-beps-project-for-discussion-at-g20-finance-ministers-meeting.htm}, accessed on November 25, 2020.
To tackle this issue, the OECD set up a project called ``Base Erosion and Profit Shifting'' (BEPS), involving over 80 countries.
See \url{http://www.oecd.org/tax/beps}, accessed on February 20, 2019.
Recent empirical studies estimate the magnitude of revenues losses from BEPS (e.g., \citealp{Dharmapala2014}; \citealp{JanskyPalansky2019}; \citealp{Beeretal2020}; \citealp{Torslovetal2020}), among which \citet{BlouinRobinson2020} caution against the possibility of overestimation due to double counting of foreign income.
The estimated magnitude of revenue losses depends on how pre-tax profits respond to corporate taxes.
\citet{HeckemeyerOveresch2017} suggest that a dominant channel of profit shifting is transfer pricing, which our model highlights.}
Our finding may justify the concern about low-tax countries attracting affiliates that receive shifted profits from high-tax countries.

The basic model is further extended to consider tax competition between two countries that differ in tax-administration efficiency. 
The main result still holds: transfer pricing leads to production agglomeration in the high-tax country with more efficient tax administration.
Contrary to existing studies telling that agglomeration generates taxable rents (\citealp{BaldwinKrugman2004}),
agglomeration in our model leads to tax base erosion.
A larger tax difference brings more opportunities to manipulate transfer prices, triggering greater tax base erosion.
To prevent this, the high-tax country is forced to lower its tax rate.
In addition, transfer pricing makes tax competition tougher and the two countries worse off by narrowing the equilibrium tax difference.
Although the governments have difficulty coordinating their tax rates,
they agree on tightening transfer-pricing regulation to achieve a Pareto improvement.

\

\noindent
{\it Relation to the literature.} \ \ \ \ 
Our main contribution to the literature on transfer pricing pioneered by \citet{Copithorne1971} and \citet{Horst1971} is to examine the impact of economic integration on the location choice of MNEs using transfer pricing.
Earlier studies point out that transfer prices are used to make affiliates competitive and to shift profits (\citealp{ElitzurMintz1996}; \citealp{SchjelderupSorgard1997}; \citealp{Zhao2000}; \citealp{Nielsenetal2003}).\footnote{For subsequent developments, see \cite{Nielsenetal2008}; \cite{ChoeMatsushima2013}; and \cite{Yao2013}.}
The former is called a strategic effect and the latter a tax manipulation effect.
These studies assume the fixed location of affiliates, unlike ours.
In our monopolistically competitive model \`{a} la \cite{DixitStiglitz1977}, optimal transfer prices themselves are not chosen strategically in the sense that they do not depend on the number of rival firms as a result of the constant elasticity of substitution between varieties.
However, our model shares the strategic aspect of transfer pricing in the sense that MNEs choose their location of plants/affiliates to make them competitive in their markets.
We allow for the flexible location of affiliates and show that it is in fact chosen strategically due to transfer pricing.

Although many studies have examined the production location choice of MNEs with profit-shifting motives, 
they focus on symmetric tax rates (e.g., \citealp{HauflerSchjelderup2000}; \citealp{Kindetal2005}; \citealp{SlemrodWilson2009}).
Some studies highlight asymmetric tax rates resulting from tax competition between symmetric/asymmetric  countries (\citealp{Stowhase2005, Stowhase2013}; \citealp{Johannesen2010}).\footnote{In many studies dealing with asymmetric tax rates, country asymmetry results from differences in market size (\citealp{Stowhase2005, Stowhase2013}).
In other tax competition models without profit-shifting MNEs, countries are assumed to be asymmetric due to inequality in public infrastructure (\citealp{Hanetal2018RIE}), the hub-and-spoke structure of jurisdictions (\citealp{JanebaOsterloh2013}; \citealp{Darbyetal2014}), and heterogeneous efficiency in tax administration (\citealp{Hanetal2014}) as in our study.}\footnote{Some studies introduce a low-tax country with no production and/or no consumption, calling it a {\it tax haven} country (\citealp{SlemrodWilson2009}; \citealp{Johannesen2010};  \citealp{Krautheimetal2011}; \citealp{Langenmayretal2015}; \citealp{Hauck2019}).
Underlying channels through which profits are shifted to tax havens include royalty payments for intangible assets (\citealp{Juraneketal2018}; \citealp{Choietal2019}) and the financial choice between debt and equity (\citealp{Fuestetal2005}; \citealp{HauflerRunkel2012}).
However, we do not consider tax havens because our main focus is on the MNEs' production location.}
\cite{Stowhase2013} studies tax competition between two unequal-sized countries and finds that the large country sets a higher tax rate while attracting a plant.
By contrast, \cite{Stowhase2005} and \cite{Johannesen2010} obtain the opposite result that low-tax countries attract more plants (or a higher capital-labor ratio) than high-tax countries.
These studies, however, do not consider trade costs, which is our primary interest.
Trade costs are important for profit-shifting patterns among MNEs because they significantly affect intra-firm trade, one of the main channels of profit shifting (\citealp{HeckemeyerOveresch2017}).

The location choice of MNEs' production plants is sometimes associated with the choice of their organizational form (\citealp{BauerLangenmayr2013}; \citealp{EggerSeidel2013}; \citealp{KeuschniggDevereux2013}).\footnote{See also \cite{AmerighiPeralta2010}; \cite{Behrensetal2014}; \cite{BondGresik2020}; and \cite{Choietal2020} for related studies on organizational choice of MNEs with profit-shifting motives.
As in the aforementioned studies, they only deal with a single MNE and/or do not consider trade costs, unlike our model.
Our companion study investigates the location of input production within the boundaries of MNEs (\citealp{KatoOkoshi2019}).}
Specifically, firms choose whether they should undertake FDI to manufacture inputs within their firms (i.e., vertical integration), or source inputs from independent suppliers (i.e., outsourcing), known as the {\it make or buy} decision.\footnote{For quantitative studies on MNEs and taxes but without transfer pricing, see \cite{Shen2018}; and \cite{Wang2020}.}
This type of FDI can be considered as vertical FDI in the sense that different stages of bringing a product on to the market are organized across borders (\citealp{AntrasHelpman2004}).
To highlight MNEs' organizational choices, whether vertical integration or outsourcing, 
these studies fix the headquarters' location, either a high-tax or low-tax country.
By contrast, we allow for MNE's location choices and examine how they are affected by economic integration, while fixing their organizational form, i.e., vertical integration only.
Furthermore, we extend the basic model to incorporate a flexible choice of organizational form and confirm the robustness of our main results (see Section 4.2).

Among studies on internationally mobile MNEs with profit-shifting motives, \cite{Peraltaetal2006}; and \cite{MaRaimondos2015} are the closest to ours in that they allow for both trade costs and asymmetric tax rates.
In tax competition over a {\it single} MNE with a plant and an affiliate,
\cite{Peraltaetal2006} show the possibility that the large, high-tax country wins the MNE's plant when trade costs are {\it high}, which is similar to but different from our findings.
Despite its higher tax rate, the large country can attract the plant by adopting a loose regulation policy, and thus, its {\it effective tax rate} is lower.
Unlike their models, we consider a {\it continuum} of MNEs, who compete in the good's market.
MNEs strategically locate their plants/affiliates such that transfer pricing contributes to price competitiveness as well as to tax savings.
This leads to a new finding in the literature, i.e., the non-monotonic impact of economic integration on MNEs' plant share, which we empirically confirm (see Section 3.3).

We also contribute to the literature on new economic geography (NEG), which examines the impact of economic integration on firm location (\citealp{Fujitaetal1999}).
To our knowledge, our study is the first to introduce transfer pricing into an NEG model, the one developed by \cite{MartinRogers1995}; and \cite{Pfluger2004}.
An important insight from NEG models is that  countries with large home markets hosts a greater share of firms than their market-size share for {\it all levels} of trade costs, except for prohibitive and zero levels.
This is known as the {\it home-market effect} (\citealp{HelpmanKrugman1985}).\footnote{To see this, consider a large and a small country, each of which initially hosts one firm.
Under a positive, but not prohibitive level of trade costs, the firm in the large country makes a greater profit by serving the larger home market without trade costs.
Thus, the firm in the small country has an incentive to relocate to the large country.}
Our model also inherits this effect in the sense that the low-tax country offers a greater profit potential for MNEs than the high-tax country.
Indeed, with high trade costs, the home-market effect dominates and the low-tax country attracts a higher plant share.
By contrast, with low trade costs, the strategic and tax-saving considerations dominate the home-market effect so that the location pattern reverses (Propositions 1 and 2 in Section 3.1).

In the analysis of tax competition using NEG models, the home-market effect allows the country, where firms are agglomerated, to set a higher tax rate without losing firms (Proposition 3 in Section 5.1; \citealp{BaldwinKrugman2004}; \citealp{BorckPfluger2006}).\footnote{See also \cite{Kindetal2000}; \cite{LudemaWooton2000};  \cite{AnderssonForslid2003}; and \cite{OttavianoYpersele2005} for earlier contributions.
Recent studies in the literature allow for heterogeneity among firms (\citealp{DaviesEckel2010}; \citealp{HauflerStahler2013}; \citealp{BaldwinOkubo2014SEA}), forward-looking behavior of governments (\citealp{Hanetal2014}; \citealp{Kato2015}).
See also \citet[Section 3.5.3]{KeenKonrad2013}.}\footnote{This result is in contrast with that of perfectly competitive models of tax competition between asymmetric countries (\citealp{Bucovetsky1991}; \citealp{Wilson1991}; \citealp{Stowhase2005}).
In these models, diminishing returns to marginal capital investment imply that smaller countries face a higher outflow of capital when raising their tax rate than larger countries, unlike NEG models.
As a result, smaller countries set a lower tax rate and achieve a higher capital-labor ratio.}
Introducing transfer pricing does not alter this location pattern, but drastically changes the implication of agglomeration.
That is, the agglomeration of production plants in the high-tax country does not bring such taxable rents, but, on the contrary, induces tax base erosion, and thus, puts downward pressure on its tax rate (Proposition 4 in Section 5.2).
If the two countries can choose the stringency of transfer-pricing regulation as well as taxes, they agree on tightening regulation in the high-tax agglomerated country and are better off (Proposition 5 in Section 5.3).\footnote{This result is in contrast with existing studies on tax competition and profit shifting with multiple policy instruments (\citealp{Peraltaetal2006}; \citealp{Hauck2019}; \citealp{HindriksNishimura2021}), where governments typically fail to coordinate any policy instruments to achieve a Pareto-improving outcome.}

The rest of the paper is organized as follows.
The next section develops the model. 
Section 3 characterizes the equilibrium plant distribution when taxes are given.
It shows how allowing for transfer pricing changes the plant distribution.
Section 4 discusses several extensions of the basic model.
Section 5 deals with tax competition between the two countries, and examines how the results change with and without profit shifting.
The final section concludes.
Formal proofs and additional robustness analyses are relegated to Online Appendix.

\

\section{Basic setting}

We consider an economy with two countries (countries 1 and 2), two goods (homogeneous and differentiated goods), and two factors of production (labor and capital).
Letting $L$ be the world population, 
there are $L_1 = s_1 L$ of population in country 1 and $L_2 = s_2 L = (1-s_1) L$ in country 2, where $s_1 \in (0,1)$ is country 1's world share.
Likewise, the amount $K$ of world capital is distributed such that country 1 (or country 2) is endowed with $K_1 = s_1 K$ (or $K_2 = s_2 K = (1-s_1)K$).
An individual in each country owns one unit of labor and two units of capital, implying that $K=2L$.
To highlight corporate tax differences, we assume away the difference in market size, i.e., $s_1=1/2$, throughout the paper except for Section 4.3.
There are two types of MNEs, one with a production plant (headquarters) in country 1 and a foreign distribution affiliate in country 2; and the other with a production plant (headquarters) in country 2 and a foreign distribution affiliate in country 1.
MNEs use labor and capital supplied by individuals.
The government in each country taxes the operating profits (i.e., sales minus labor/input costs) of plants and affiliates there.
We interpret capital as equity and assume that capital costs are non-deductible, while labor costs are deductible.
This assumption is commonly adopted in the public finance literature mentioned in Introduction and is in line with the tax codes of OECD countries.\footnote{The Glossary of Tax Terms of the OECD website defines ``DIVIDENDS'' as follows: ``A payment by a corporation to shareholders [$\cdots$]. Most corporations receive no deduction for it.'' See: \url{https://www.oecd.org/ctp/glossaryoftaxterms.htm}, accessed on December 19, 2021.}
In this and the next sections, we fix the tax rates of countries and assume that country 1's tax rate is higher than that of country 2, $t_1 > t_2$, without loss of generality.
The key variables are listed in Appendix 1.

The timing of actions is as follows.
First, each MNE chooses the respective country in which to locate a production plant and a foreign distribution affiliate, thereby endogenously determining the share of plants $n_1$.
Second, the MNE chooses transfer prices, $g_i$.
Third, production plants and distribution affiliates set selling prices, $p_{ij}$.
Finally, production and consumption take place.
We solve the game backward.
For convenience, we refer to the results with fixed capital allocation as a {\it short-run equilibrium} and refer to the results in the endogenous case as a {\it long-run equilibrium}.

\

\noindent
{\it Consumers.} \ \ \ \ Following \cite{Pfluger2004}, each consumer has an identical quasi-linear utility function with a constant-elasticity-of-substitution (CES) subutility.
Consumers in country 1 solve the following maximization problem:
\begin{align*}
&\max_{\tilde{q}_{11}(\omega), \tilde{q}_{21}(\omega), q_1^O} \ u_1 = \mu \ln Q_1 + q_1^O, \\
\
&\text{where} \ \ Q_1 \equiv \left[ \sum_{i = 1}^2 \int_{\omega \in \Omega_i} \tilde{q}_{i1}(\omega)^{\frac{\sigma-1}{\sigma}}d\omega \right]^{\frac{\sigma}{\sigma-1}},
\end{align*}
subject to the budget constraint:
\begin{align*}
\sum_{i = 1}^2 \int_{\omega \in \Omega_i} p_{i1}(\omega) \tilde{q}_{i1}(\omega) d\omega +  q_1^O = w_1 + \overline{q}_1^O.
\end{align*}
$\mu > 0$ captures the intensity of the preference for the differentiated goods.
$q_1^O$ and $\overline{q}_1^O$ are respectively the individual demand for the homogeneous good and its initial endowment.
We assume that $\overline{q}_1^O$ is large enough for the homogeneous good to be consumed.
$w_1$ is the wage rate.
$\tilde{q}_{i1}(\omega)$ is the individual demand from consumers in country 1 for the variety $\omega \in \Omega_i$, where $\Omega_i$ is the set of varieties produced in country $i \in \{1,2 \}$.
$Q_1$ is the CES aggregator of differentiated varieties, with $\sigma>1$ being the elasticity of substitution over them.

Solving the foregoing problem gives the aggregate demand for the variety $\omega$ produced in country $i \in \{ 1, 2\}$ and consumed in country 1:
\begin{align*}
&q_{i1}(\omega) \equiv L_1 \tilde{q}_{i1}(\omega) = \left( \frac{p_{i1}(\omega)}{P_1} \right)^{-\sigma} \frac{\mu L_1}{P_1}, \tag{1} \\
\
&\text{where} \ \ P_1 \equiv \left[ \sum_{i = 1}^2 \int_{\omega \in \Omega_i} p_{i1}(\omega)^{1-\sigma}d\omega \right]^{\frac{1}{1-\sigma}}.
\end{align*}
$P_1$ is the price index of the varieties.
Although we will mainly present the results for country 1 in the following, analogous expressions hold for country 2.
As firms are symmetric, we suppress the variety index $\omega$ for notational brevity.

\

\noindent
{\it Homogeneous good sector.} \ \ \ \ The homogeneous good sector uses a constant-returns-to-scale technology.
That is, one unit of labor produces one unit of the good.
The technology leads to perfect competition, making the good's price equal to its production cost, or the wage rate.
Letting $w_i$ be the wage rate of country $i \in \{ 1, 2\}$, 
the costless trade of the homogeneous good equalizes the wage rates between countries, i.e., $w_1=w_2$.\footnote{We assume the costless trade of the homogeneous good to highlight the role of differentiated sector while eliminating the complicated general-equilibrium channel of wages.
This is common in the NEG models, but is not totally innocuous. See, for example, \citet[Chapter 7]{Fujitaetal1999} for more on this.}
We choose the good as the num{\'e}raire such that $w_1=w_2=1$.

\

\noindent
{\it Differentiated goods sector.} \ \ \ \ The differentiated goods sector uses an increasing-returns-to-scale technology.
Each MNE needs one unit of capital for a production plant serving as the headquarters in one country and another unit for a distribution affiliate in the other country.\footnote{Similar specifications in the context of transfer pricing can be found in \cite{Kindetal2005}; and \cite{Matsui2012}, although they fix the location of plants and affiliates.}
Once established, the plant needs $a$ units of labor to produce one unit of variety.
Since the total amount of capital in the world is $K = 2L$, the mass of (the headquarters of) MNEs in the world is $K/2=L$.\footnote{All qualitative results in this study do not depend on the size of $L$. It matters only for the level of the equilibrium tax rate in Section 5 and the level of tax revenues in Online Appendix E.}
We denote the mass of production plants located in country 1 by $N_1 = n_1 L$ and that in country 2 by $N_2 = n_2 L = (1-n_1)L$, where $n_1 \in [0, 1]$.
Each MNE owns both a plant and an affiliate.
Thus, country 1 hosts a mass $N_1$ of plants and $N_2$ of affiliates, while country 2 hosts a mass $N_2$ of plants and $N_1$ of affiliates.

Consider an MNE with a production plant (i.e., headquarters) in country 1.
The plant produces quantities $q_{11}$ using $a q_{11}$ units of labor and sells them at a price $p_{11}$ to home consumers.
In addition, it produces quantities $q_{12}$ and exports them at a transfer price $g_1$ to its distribution affiliate in country 2.
When exporting, due to iceberg trade costs $\tau>1$, a $(\tau-1)/\tau$ fraction of quantities melts away. Thus, the plant has to produce $\tau$ units to deliver one unit to the affiliate.
The affiliate sells the imported goods to consumers in country 2 at a price $p_{12}$.

MNEs are assumed to have decentralized decision making following previous studies on transfer pricing (\citealp{Zhao2000}; \citealp{Nielsenetal2003, Nielsenetal2008}; \citealp{Kindetal2005}).
In other words, the production plant (i.e., the headquarters) of the MNE sets the transfer price to maximize global post-tax profits, while the foreign affiliate sets the retail price to maximize its own profits.
In practice, it is sensible to delegate decisions to local managers who are familiar with their local business environments.
In many cases, a company's acquisition of a rival often involves the latter receiving divisional autonomy (e.g., Volkswagen's
acquisition of Audi, Ford's acquisition of Volvo, and GM's acquisition of Saab).\footnote{See \cite{Ziss2007} for more on this issue.}
We examine the case of centralized decision making in Online Appendix H and confirm the robustness of our results. 

\

\subsection{Short-run equilibrium}

Here, we derive the optimal prices given the location of plants and affiliates (see Online Appendix A for detailed derivations).
The initial share of production plants is assumed to be equal to the capital endowment share, i.e., $n_1 = s_1 = 1/2$.\footnote{As we shall see, the allocation of plants does not affect optimal selling prices (i.e., transfer prices).}
The MNE with production in country 1 makes profits from a home plant and a foreign distribution affiliate in country 2.
The pre-tax operating profits of the plant, $\pi_{11}$, and those of the affiliate, $\pi_{12}$, are respectively,
\begin{align*}
&\pi_{11} = \underbrace{(p_{11} -a)q_{11}}_{\text{Domestic profits}} 
+\underbrace{(g_1 -\tau a)q_{12}}_{\text{Shifted profits}}
-\underbrace{\delta|g_1 -\tau a|q_{12}}_{\text{Concealment cost}}, \\
\
&\pi_{12} = \underbrace{(p_{12} -g_1)q_{12}}_{\text{Foreign profits}} \\
\
&\ \ \ \ = (p_{12} -\tau a)q_{12} \underbrace{-(g_1 -\tau a)q_{12}}_{\text{Shifted profits}},
\end{align*}
where $q_{11}$ is given by Eq. (1); $q_{12}$ is defined analogously; and $a$ is the unit labor requirement.
The second term in $\pi_{11}$ represents the profits from intra-firm trade subject to trade costs $\tau$.
The MNE may choose the transfer price differently than the true marginal cost: $g_i \neq \tau a$.
For profits to move from high-tax country 1 to low-tax country 2, the second term must be negative: $(g_1 -\tau a)q_{12} < 0$ or $g_1 < \tau a$, which we will see shortly.
This term captures profit shifting within MNEs, appearing in $\pi_{12}$ with the opposite sign.
The third term in $\pi_{11}$ is concealment costs associated with the deviation of the transfer price from the true marginal cost (\citealp{HauflerSchjelderup2000}; \citealp{Kindetal2005}).
As the deviation is larger, it is more costly for MNEs to conceal the transfer pricing activity from tax authorities.
A high $\delta$ makes profit shifting more costly, implying that $\delta$ can be interpreted as the stringency of transfer price regulation.

In the third stage of the game, the production plant and the foreign affiliate choose their prices to maximize their own profits.
The optimal prices are
\begin{align*}
&p_{11} = \frac{\sigma a}{\sigma-1}, \ \ \ \ p_{12} = \frac{\sigma g_1}{\sigma-1}. \tag{2}
\end{align*}

At the second stage, the MNE with a plant in the high-tax country 1 sets the transfer price to maximize the following global post-tax profits $\Pi_1$:
\begin{align*}
\Pi_1 \equiv (1-t_1)\pi_{11} + (1-t_2)\pi_{12} - 2R_1, \tag{3}
\end{align*}
where $t_i \in [0,1]$ is the tax rate of country $i \in \{ 1, 2\}$ and $R_1$ is the reward to capital invested in the MNE.
Again, note that labor/input costs are deductible, while capital costs are non-deductible.
Suppose that the MNE with production in country 1 tries to shift profits of its home plant to its foreign affiliate in the low-tax country 2, i.e., $(g_1-\tau a)q_{12} < 0$, where the concealment cost is given by $\delta| g_1 -\tau a|q_{12} = -\delta (g_1 -\tau a)q_{12} > 0$.
It chooses the transfer price to maximize the post-tax profit (Eq. (3)):
\begin{align*}
&g_1 = \frac{(1+\delta) \sigma \tau a}{\sigma - \Delta{t}_1 + \delta (\sigma-1)}, \ \ \ \ \text{where} \ \ \Delta{t}_1 \equiv \frac{t_2-t_1}{1-t_1} < 0, \tag{4-1}
\end{align*}
where $g_1$ decreases with $t_1$ and increases with $t_2$.
This implies that a greater tax difference leads to a more aggressive transfer-pricing behavior.\footnote{The result that a larger tax difference leads to a lower export price from the high-tax to low-tax country is in line with empirical findings by \cite{Clausing2003}.}
Similarly, supposing that the MNE with production in country 2 tries to shift profits from its foreign affiliate in country 1 back to its home plant,
it incurs the concealment cost of $\delta |g_2 -\tau a|q_{21} = \delta (g_2 -\tau a)q_{21} > 0$.  
The optimal transfer price maximizing the MNE's post-tax profit is
\begin{align*}
&g_2 = \frac{(1-\delta)\sigma \tau a}{\sigma - \Delta{t}_2 -\delta(\sigma-1)}, \ \ \ \ \text{where} \ \ \Delta{t}_2 \equiv \frac{t_1-t_2}{1-t_2} > 0, \tag{4-2}
\end{align*}
where $g_2$ decreases with $t_2$ and increases with $t_1$.
The optimal transfer prices $g_i$ do not depend on the plant share $n_i$ because of the constant elasticity of substitution $\sigma$, implying that $g_i$s entail only tax-saving motives, not strategic ones.
As will be clear in the next section, however, MNEs make a strategic location choice such that they use $g_i$s to make their affiliates competitive.

For the optimal transfer prices to be consistent with the direction of profit shifting, the optimal transfer price from country 1 to 2 (or from country 2 to 1) must be set lower (or higher) than the true marginal cost:
\begin{align*}
&g_1 < \tau a \ \to \ \delta < \frac{t_1-t_2}{1-t_1}, \\
\
&g_2 > \tau a \ \to \ \delta < \frac{t_1-t_2}{1-t_2}.
\end{align*}
These conditions reduce to $\delta < (t_1-t_2)/(1-t_2) = \overline{\delta}$, which also satisfies the second-order condition for maximization ($\delta < 1$).
Under $\delta < \overline{\delta}$, profits (net of concealment costs) shifted from country 1 to 2 are negative: $(1+\delta)(g_1-\tau a)q_{12}<0$.
By contrast, those shifted from country 2 to 1 are positive: $(1-\delta)(g_2-\tau a)q_{21}>0$.
If the tax difference is too large, the total pre-tax profits of the plant in country 1 could be negative: $\pi_{11}<0$ (see Online Appendix B for details). 
To exclude this possibility, we further assume $(t_2 <) t_1 < 1/2$, which is plausible considering the highest corporate tax rate being $0.4076$ in 23 OECD countries in 2010 to 2016 (from 2010 to 2012 in Japan).

Using the demand function (Eq. (1)), and optimal prices (Eqs. (2) and (4)), we rearrange the post-tax profit (Eq. (3)) as
\begin{align*}
&\Pi_1 = (1-t_1)\pi_{11} + (1-t_2)\pi_{12} -2R_1 \\
\
&\ \ \ = (1-t_1) \biggl[ \underbrace{\frac{\mu L_1}{\sigma(N_1 +\phi \gamma_2 N_2)}}_{\text{Domestic profits}}  
+ \underbrace{\frac{(\sigma-1)(\Delta{t}_1 + \delta)}{\sigma} \cdot \frac{\phi \gamma_1 \mu L_2}{\sigma(\phi \gamma_1 N_1 +N_2)}}_{\text{Shifted profits (net of concealment cost)} \ < 0} \biggr] \\
\
&\ \ \ \ \ \ \ \ + (1-t_2) \cdot \underbrace{\frac{\phi \gamma_1 \mu L_2}{\sigma(\phi \gamma_1 N_1 +N_2)}}_{\text{Foreign profits}} - 2R_1, \tag{5-1} \\
\
&\Pi_2 = (1-t_1)\pi_{21} + (1-t_2)\pi_{22} -2R_2 \\
\
&\ \ \ = (1-t_1) \cdot \underbrace{\frac{\phi \gamma_2 \mu L_1}{\sigma(N_1 +\phi \gamma_2 N_2)}}_{\text{Foreign profits}} \\
\
&\ \ \ \ \ \ \ \ + (1-t_2) \biggl[ \underbrace{\frac{\mu L_2}{\sigma(\phi \gamma_1 N_1 + N_2)}}_{\text{Domestic profits}} 
+ \underbrace{\frac{(\sigma-1)(\Delta{t}_2 - \delta)}{\sigma} \cdot \frac{\phi \gamma_2 \mu L_1}{\sigma( N_1 +\phi \gamma_2 N_2)}}_{\text{Shifted profits (net of concealment cost)} \ > 0} \biggr] -2R_2, \tag{5-2} \\
\
&\text{where} \ \ \phi \equiv \tau^{1-\sigma}, \ \ \ \ \Delta{t}_i \equiv \frac{t_j -t_i}{1-t_i}, \ \ \ \ i \neq j \in \{ 1,2\}, \\
\
&\ \ \ \ \ \ \ \ \ \gamma_1 \equiv \left( \frac{\sigma (1+\delta)}{\sigma- \Delta{t}_1 +\delta(\sigma-1)} \right)^{1-\sigma}, \ \ \ \ \gamma_2 \equiv \left( \frac{\sigma (1-\delta)}{\sigma- \Delta{t}_2 -\delta(\sigma-1)} \right)^{1-\sigma}.
\end{align*}
The first and second terms in the square brackets in $\Pi_1$ and $\Pi_2$ are respectively the profit from the domestic market and the profit shifted through transfer pricing.
$\phi = \tau^{1-\sigma} \in [0,1]$ is an inverse measure of trade costs, or the openness of trade.
$\phi=0$ (i.e., $\tau=\infty$) corresponds to a prohibitively high level of trade costs, while $\phi=1$ (i.e., $\tau=1$) indicates zero trade costs.

We assume, as in standard NEG models, the free entry and exit of potential MNEs (or equivalently the arbitrage behavior of capital owners), so that excess profits are driven to zero.
This zero-profit condition implies that for given MNEs' locations, the return to capital, $R_i$, is determined at the point where $\Pi_i = 0$ holds.
In the short-run equilibrium, where capital is immobile, the return to capital in general differs between countries.
The capital-return differential generates a relocation incentive which guides us to analyze the long-run equilibrium, where capital is mobile.

\

\section{Long-run equilibrium}

To highlight the role of tax difference, we assume that the two countries are of equal size ($s_1=1/2$).
When the difference in capital returns is positive, $R_1-R_2 > 0$ (or negative, $R_1 - R_2 < 0$), capital owners invest in the MNE with production in country 1 (or in country 2).
In the long-run equilibrium, the return differential is zero,  with no capital owners changing their investment behavior.
This also means that no MNEs are willing to change the location of their plants/affiliates.
By solving the long-run equilibrium condition ($R_1 -R_2 =0$) for the share of production plants in country 1,
we obtain interior equilibria $n_1 \in (0,1)$.
If $R_1 -R_2 =0$ does not have interior solutions, then we obtain corner equilibria in which all multinational production takes place in one country, i.e., $n_1 \in \{ 0, 1\}$.

\subsection{Plant distribution and the non-monotonic impact of economic integration}

The location incentives of MNEs depend on the ease of intra-firm trade, which is subject to trade costs.
A low trade openness $\phi$ does not allow for much intra-firm trade, leaving little room for profit shifting.\footnote{In Eqs. (5-1) and (5-2), the profits from intra-firm trade and those from the foreign affiliate disappear if $\phi=0$.} 
As MNEs earn profits mostly from home production plants, they prefer to locate them in the low-tax country 2.
The concentration of production in country 2 results from the well-known home-market effect, which is commonly observed in NEG models (\citealp{HelpmanKrugman1985}).
We note that there is a small, but positive share of plants in the high-tax country 1, i.e., $n_1|_{\phi=0} \in (0, 1/2)$.
Since competition in the domestic market works as a dispersion force,
the corner distribution where all plants are in the low-tax country 2 cannot be an equilibrium, i.e., $n_1|_{\phi=0} \neq 0$. 

By contrast, a high $\phi$ allows MNEs to fully engage in intra-firm trade, increasing the effectiveness of profit shifting through transfer pricing.\footnote{We can confirm that given the plant share $n_1$, the shifted profit increases with $\phi$; that is, $\partial (|g_i -\tau a|q_{ij})/ \partial \phi > 0$ for $i \neq j \in \{ 1, 2\}$.}
Transfer pricing does not just shift profits between home plants and foreign affiliates, but also affects the competitiveness of foreign affiliates.
As we showed, MNEs with production in the high-tax country 1 set a low transfer price to shift profits to their affiliates in the low-tax country 2 (see Eq. (4-1)).
Due to the low input cost, affiliates can sell at a low price and become competitive against local plants.
Conversely, MNEs with production in country 2 set a high transfer price (see Eq. (4-2)), which makes their affiliates in country 1 less competitive against local plants.
MNEs with production in country 1 have a competitive advantage in both domestic and foreign markets against MNEs with production in country 2.
Therefore, MNEs prefer to locate production in the high-tax country 1 so that transfer pricing makes their affiliates competitive.
In fact, if $\phi$ is sufficiently high such that $\phi \ge \phi^S$, which we call an agglomeration threshold, all production plants are located in country 1.\footnote{The agglomeration threshold is called the sustain point in the literature in the sense that full agglomeration is sustainable when trade openness is higher than this point (\citealp{Fujitaetal1999}).
We can verify that $\phi^S$ decreases with $t_1-t_2$ (see Online Appendix D).
A larger tax difference offers more room for profit shifting, and thus, leads to more aggressive transfer pricing (very low $g_1$ or very high $g_2$).
This strengthens the competitiveness of MNEs with production in country 1 against those with production in country 2 in both domestic and foreign markets.
Consequently, a larger tax difference lowers $\phi^S$, making more likely full production agglomeration in country 1.}

Assuming away transfer-pricing regulation ($\delta=0$), we can formally prove the following proposition by applying Taylor approximations at zero tax difference.

\

\noindent
{\bf Proposition 1 (Plant distribution).} \ \ {\it Assume that two countries are of equal size ($s_1=1/2$), country 1 has a higher corporate tax rate than country 2 ($t_2 < t_1 < 1/2$), the tax difference is small enough ($t_1 -t_2 \approx 0$), and there is no transfer-pricing regulation ($\delta = 0$).
There exist two threshold values of trade openness, $\phi^{\dagger} < \phi^S$,  such that the following holds:
\begin{itemize}
\item[(i).] If $\phi \in [0, \phi^{\dagger})$, the high-tax country 1 hosts a smaller share of plants than the low-tax country 2, i.e., $n_1 \in [0, 1/2)$.
\item[(ii).] If $\phi \in [\phi^{\dagger}, \phi^S)$, the high-tax country 1 hosts an equal or a greater share of plants but does not attract all plants, i.e., $n_1 \in [1/2, 1)$.
\item[(iii).] If $\phi \in [\phi^S, 1]$, the high-tax country 1 attracts all plants, i.e., $n_1=1$.
\end{itemize}
}

\

See Online Appendix C for the proof, where we also numerically check that Proposition 1 holds when the tax difference is not close to zero under a plausible range of parameter values.
We do not report here the lengthy expressions of $\phi^{\dagger}$ and $\phi^{S}$, which are implicitly defined as $\phi^{\dagger} \equiv \text{arg}_{\phi} \{ n_1 = 1/2\}$ and $\phi^S \equiv \min \text{arg}_{\phi} \{ n_1 = 1\}$, respectively.

Notably, full production agglomeration in country 1 occurs even with completely free trade ($\phi=1$).
In the case without transfer pricing, MNEs are indifferent to the location of plants at $\phi=1$.
The selling price for the foreign market is equal to that for the home market, i.e., $p_{ij} = p_{ii} = \sigma a/(\sigma-1)$, so that MNEs make the same profits from the two markets.
They cannot avoid high taxes in country 1 by changing the location of plants/affiliates, and thus, do not have a strong location preference.
In the transfer-pricing case, however, the selling prices for the two markets are not equalized even at $\phi=1$, i.e., $p_{ij} \neq p_{ii}$, because they depend on taxes (see Eqs. (4-1) and (4-2)).
To fully utilize transfer pricing to save taxes and to enhance competitiveness, MNEs have a strong preference for production location even at $\phi=1$.

The effect of trade openness on the world distribution of production is illustrated in Figure 2.
Figure 2(a) shows a representative pattern of long-run equilibrium plant share for different levels of $\phi$ (solid curve), along with the long-run equilibrium plant share in the case without transfer pricing (dashed curve).\footnote{Parameter values for all figures are also summarized in Appendix 4.
We set the elasticity of substitution $\sigma$ to five, which is the lower bound of the range estimated by \cite{LaiTrefler2002} and is close to the median estimate by \cite{BrodaWeinstein2006}.
We set tax rates to $(t_1, t_2)=(0.3, 0.2)$, which seems plausible as the average tax rate is $0.274$ and the average tax difference is $0.077$ in our sample of 23 OECD countries from 2008 to 2016.
In the special case where the tax difference is extremely large or $\sigma$ is extremely low, it is possible that as openness is higher, full production agglomeration in country 2 occurs before full production agglomeration in country 1 is achieved.
However, our qualitative results remain unchanged in this special case.
See Online Appendix D for details.}
Figure 2(b) depicts an enlarged view of Figure 2(a) to highlight the threshold values.
As $\phi$ increases from zero, the share of plants in the high-tax country 1 decreases in both cases with and without profit shifting.
When low openness prevents exporting, MNEs make profits mostly from their home plants, and thus, prefer to locate them in the low-tax country 2.
Along with a further increase in $\phi$ from $\phi^{\#}$, however, the mass of plants in the high-tax country 1 increases in the case with profit shifting, whereas it continues to decrease  in the case without profit shifting.
Sufficiently high openness expands intra-firm trade, and thus, increases the opportunities for profit shifting, leading to a sharp contrast in location patterns.

\

\begingroup
\begin{center}
\includegraphics[scale=0.8]{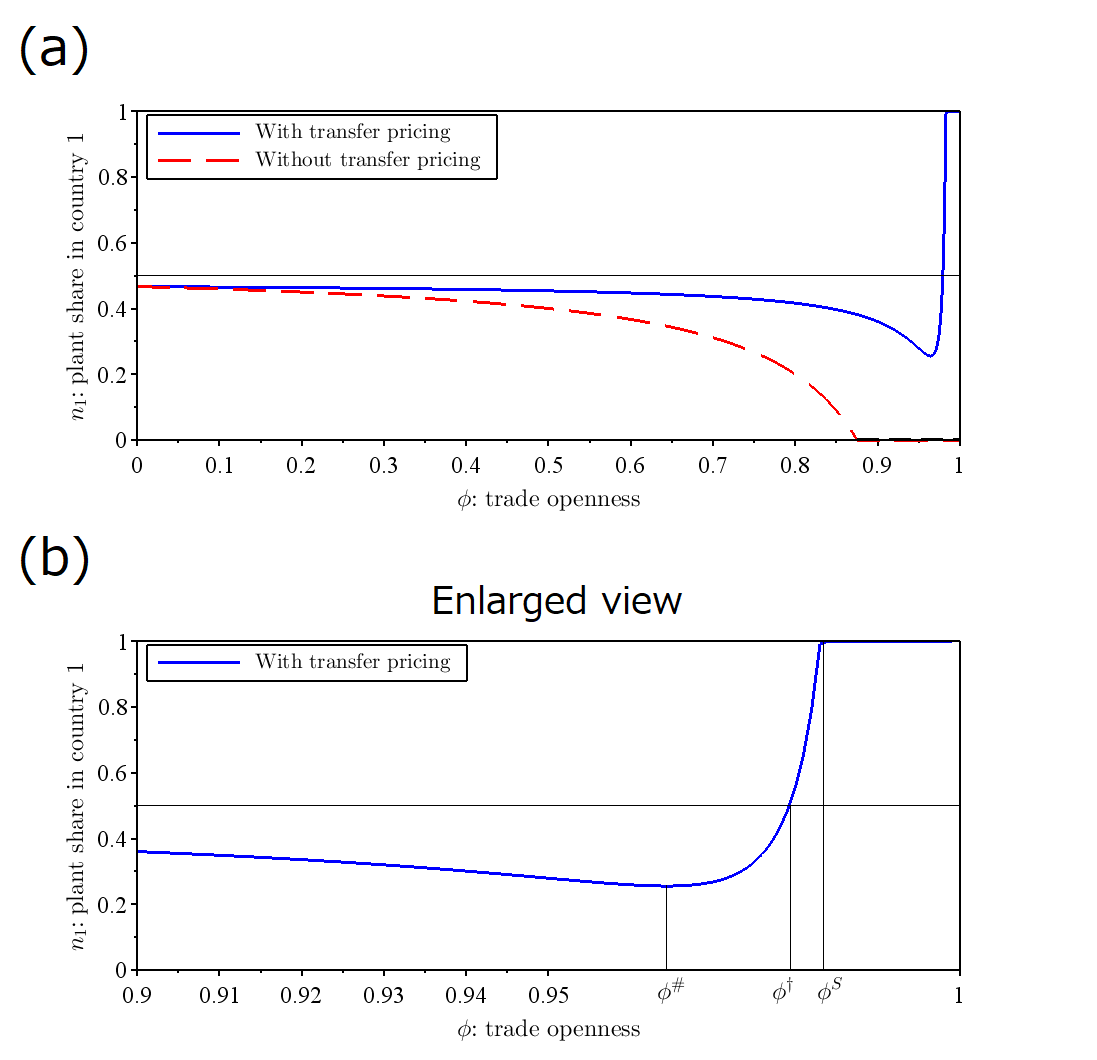} \\
Figure 2. \ Share of production plants in the high-tax country 1 \\
{\footnotesize {\it Notes}: \ Parameter values are $\sigma=5$; $t_1=0.3$; $t_2=0.2$; $\delta=0$; $s_1=0.5$.}
\end{center}
\endgroup

\

This finding is summarized in the following proposition:

\

\noindent
{\bf Proposition 2 (Non-monotonic impact of economic integration).} \ \ {\it Under the same assumptions as in Proposition 1, the impact of economic integration on the plant share in the high-tax country 1, $n_1$, is non-monotonic.
That is, a rise in trade openness first decreases and then increases the plant share, i.e., $dn_1/d\phi \leq 0$ for $\phi \in [0, \phi^{\#})$ and $dn_1/d\phi \ge 0$ for $\phi \in (\phi^{\#}, 1]$.}

\

\noindent
See Online Appendix D for the proof, where we also numerically check that Proposition 2 holds when the tax difference is not close to zero under a plausible range of parameter values.

\

\noindent
{\it Tax revenues.} \ \ \ \ As a result of transferring profits taxable in the high-tax country 1 to the low-tax country 2, introducing transfer pricing weakly reduces (or weakly raises) tax revenues in country 1 (or country 2).
We formally show this in Online Appendix E.

\

\subsection{Empirical evidence}

Proposition 2 provides the following empirical implication.
Given the tax difference between two countries, there is a non-monotonic effect of bilateral trade openness on the share of manufacturing affiliates out of those in all sectors.
To empirically test this prediction, we can think of the following regression:
\begin{align*}
&\text{(Manufacturing-affiliate share)}_{h, s, t} = \beta_1 \Delta TAX_{h, s, t} \cdot \phi_{h, s} + \beta_2 \Delta TAX_{h, s, t} \cdot \phi_{h, s}^2  + \bm{x_{h, s, t}' \beta} + \varepsilon_{h, s, t},
\end{align*}
where the variables are defined as 
\begin{align*}
&\text{(Manufacturing-affiliates share)}_{h, s, t}: \\
\
&\ \ \text{the share of manufacturing affiliates out of all affiliates in country $h$ from country $s$ in time $t$}, \\
\
&\Delta TAX_{h,s,t} = TAX_{h, t} - TAX_{s, t}: \text{corporate tax rate difference between $h$ and $s$ in time $t$}, \\
\
&\phi_{h, s}: \text{bilateral trade openness}, \\
\
&\bm{x}_{h, s, t}: \text{vector of control variables}, \\
\
&\varepsilon_{h, s, t}: \text{error term},
\end{align*}
noting that with an abuse of notation, we here use $t$ for the time subscript.
Supposing that the host country $h$ sets a higher tax rate than the source country $s$, i.e., $\Delta TAX_{h, s, t} > 0$, our theory predicts
\begin{align*}
\frac{\partial (\text{Manufacturing-affiliate share})_{h,s,t} }{ \partial \phi_{h, s} } = \underbrace{\Delta TAX_{h,s,t}}_{>0} \cdot (\beta_1 + \beta_2 \underbrace{\phi_{h,s}}_{>0} )
\begin{cases}
<0 \ \ \text{if} \ \phi_{h, s} < - \beta_1/\beta_2 \\
>0 \ \ \text{if} \ \phi_{h, s} > - \beta_1/\beta_2
\end{cases},
\end{align*}
which states that an increase in trade openness has a negative (or positive) effect on the manufacturing-affiliate share in country $h$ coming from country $s$ if trade openness is low (or high).
The sign of the derivative flips in the case of $\Delta TAX_{h, s, t} < 0$.
Because $\phi_{h, s}$ takes positive values, the theory-consistent signs are $\beta_1<0$ and $\beta_2>0$.

We test this using the same affiliate data as those used in Figure 1.
To construct bilateral trade openness, $\phi_{h,s} \in (0, 1)$, we take the inverse of the log of the bilateral geodesic distance constructed by \cite{MayerZignago2011}.
In addition to the two interaction terms of our interest, control variables ($\bm{x}_{h, s, t}$) include a simple host-source tax difference, $\Delta TAX_{h,s,t}$; a host-source difference in labor costs, $\Delta (\text{Labor costs})_{h,s,t}$; a host-source difference in productivity, $\Delta (\text{Productivity})_{h,s,t}$; a host-source difference in index capturing pro-business policies, $\Delta (\text{Pro business})_{h,s,t}$; country-year dummy; a source country-year dummy; and a year dummy, following the empirical literature on the determinants of FDI (e.g., \citealp{Eggeretal2009}; \citealp{Choietal2021}).\footnote{We cannot include variables varying only in host/source country and year such as host/source country GDP because their effects are absorbed by the host/source country-year dummy.}
In addition to the statutory tax rate, as a robustness check, we use the effective average tax rate (\citealp{DevereuxGriffith1999}), which takes care of forward-looking behavior of firms. 
The sample is an unbalanced panel of 23 OECD countries, covering the period 2008--2016.
We exclude observations with missing values and/or zero affiliates from the sample.
See Appendix 2 for details on the descriptions and sources of variables and Table A1 for summary statistics.

One may be concerned about that corporate tax policies in a country are influenced by affiliate activities there.
We address this potential endogeneity by following \cite{DaRinetal2010, DaRinetal2011}.\footnote{When estimating the effect of corporate tax rates on firm entry, \cite{DaRinetal2010, DaRinetal2011} use political measures such as the degree of political stability and election dates as IVs for corporate tax rates.}
To construct an instrumental variable (IV) for the host-source tax difference, we argue that the characteristics of a country's political system are important determinants of corporate taxation and other business-related policies, which in turn affect FDI flows.
For example, a more impartial electoral process strengthens the transparency of policymaking and helps the government gain public trust, making changing (or perhaps raising) taxes easier.
The impartial electoral process would not have a direct but indirect effect on FDI flows only through corporate taxation and other policies such as business laws/regulations and infrastructure investment.
The difference in political processes would be a valid IV for the corporate tax difference, once pro-business policies other than corporate taxes are appropriately controlled for.

Specifically, as a measure of political processes, we use a summary index of government accountability constructed by the World Governance Indicators of the World Bank.
This accountability index is the mean of 21 policy indices reported in the Institutional Profiles Database of the CEPII.
The sub-indices measure the freedom of elections at the national level, the reliability of the state budget, and the freedom of association.
We take the host-source difference of the index ($\Delta (\text{Accountability})_{h,s,t}$) and use it as an IV for the corporate tax difference.\footnote{One may wonder  countries with large market size tend to have both high corporate taxes and accountable governments.
This is not the case in our sample.
Although the log of GDP is highly correlated with the corporate tax rate ($0.75$), it has little correlation with our accountability measure ($0.047$).}
To control for other pro-business policies than corporate taxes, which affect the share of manufacturing affiliates, we include the host-source difference in pro-business policies ($\Delta (\text{Pro business})_{h,s,t}$) as a control variable.
This is calculated from the ``Index of Economic Freedom'' published by the Heritage Foundation.

The regression results are summarized in Table 1.
In columns (1) and (2), the statutory tax rates are used.
In columns (3) and (4), the effective average tax rates are used.
In columns (2) and (4), the tax difference and its interactions with $\phi_{h,s}$ and $\phi_{h,s}^2$ are instrumented with the difference in accountability and its interactions with $\phi_{h,s}$ and $\phi_{h,s}^2$.
The high values of the first-stage Cragg-Donald $F$ statistic in columns (2) and (4) suggest that our IVs are not weak.
We can also confirm that the turning point of trade openness is actually between zero and one (e.g., in column (1): $\phi^{\#} = 0.715/(2 \cdot 2.630) \simeq 0.136$) and its equivalent geodesic distance is around 1560 km ($0.136 \simeq 1/\log (1560)$), roughly equal to the one between Germany and Spain (1479.3 km).
As a further robustness check, we use a trade-cost measure calculated {\`a} la \cite{HeadRies2001}; and \cite{Novy2013} to construct an alternative trade-openness measure.
Estimation results are unchanged: see Tables A2 and A3 in Appendix 3.

\pagebreak

\begingroup
\begin{center}
Table 1 \\
Non-monotonic effect of economic integration on multinational production \\
{\it Dependent variable: \ (foreign manufacturing affiliates)$/$(foreign affiliates in all sectors)} \\
\vspace{0.3cm}
\begin{tabular}{ l  c  c  c  c  c }
\hline \hline
	 & \multicolumn{2}{c}{Statutory tax rate} &  & \multicolumn{2}{c}{Effective average tax rate} \\ \cline{2-3} \cline{5-6}
	& OLS & 2SLS &  & OLS & 2SLS \  \\ \cline{2-6}
	 & (1) & (2) &  & (3) & (4) \\ \hline
  $\Delta TAX_{h ,s, t} \cdot \phi_{h, s}$ & $-$0.715*** & $-$1.477* & & $-$0.864*** & $-$1.804*  \\ 
  & (0.218) & (0.886) & & (0.232) & (1.027)  \\ 
  $\Delta TAX_{h ,s, t} \cdot \phi_{h, s}^2$ & 2.630*** & 6.793** & & 3.191*** & 8.083**  \\ 
  & (0.789) & (3.254) & & (0.848) & (3.857)  \\ [0.3cm]
  $\Delta TAX_{h ,s, t}$ & 0.034 & 0.060 & & 0.045 & 0.083  \\ 
  & (0.127) & (0.166) & & (0.117) & (0.157)  \\ 
  $\Delta (\text{Labor cost})_{h ,s, t}$ & 0.001 & 0.006 & & 0.002 & 0.007 \\ 
  & (0.009) & (0.010) & & (0.017) & (0.019) \\ 
  $\Delta (\text{Productivity})_{h ,s, t}$ & 0.008 & 0.010 & & 0.009 & 0.010 \\ 
  & (0.015) & (0.017) & & (0.014) & (0.016) \\ 
  $\Delta (\text{Pro business})_{h ,s, t}$  & $-$0.004 & $-$0.007 & & $-$0.004 & $-$0.007 \\ 
  & (0.047) & (0.055) & & (0.042) & (0.047) \\ \hline
	Host country--year dummy & $\checkmark$ & $\checkmark$ & & $\checkmark$ & $\checkmark$ \\
	Source country--year dummy & $\checkmark$ & $\checkmark$ & & $\checkmark$ & $\checkmark$ \\ 
	Year dummy & $\checkmark$ & $\checkmark$ & & $\checkmark$ & $\checkmark$  \\ 
	Cragg--Donald $F$ &  & 20.21 &  &  & 25.98 \\ 
	Observations & 1,833 & 1,825 &  & 1,833 & 1,825 \\
	$R^2$ & 0.507 & 0.422 &  & 0.507 & 0.422 \\ \hline \hline
\end{tabular} \\
\begin{flushleft} {\footnotesize
\textit{Notes}: The dependent variable is the share of a source country's manufacturing affiliates in a host country out of the source's affiliates in all sectors in the host in a year.
Standard errors clustered at the host country-year level are in parentheses.
We use the host-source difference in government accountability ($\Delta (\text{Accountability})_{h,s,t}$) as the IV for $\Delta TAX_{h,s,t}$.
The interaction terms between a variable and $\Delta TAX_{h,s,t}$ are also instrumented by the interaction terms between the variable and $\Delta (\text{Accountability})_{h,s,t}$. \\
***Significant at the 1$\%$ level; **Significant at the 5$\%$ level; *Significant at the 10$\%$ level.}
\end{flushleft}
\end{center}
\endgroup


\pagebreak

\

Supporting evidence can also be found in \cite{Overesch2009} and \cite{Goldbachetal2019}.
Using firm-level panel data of German inward FDI from 1996 to 2005, \cite{Overesch2009} finds that domestic investment by foreign affiliates located in Germany increases as the corporate tax difference between high-tax Germany and their source country is higher.
\cite{Goldbachetal2019} also employ similar data of German outward FDI and find that the complementarity between domestic and foreign investment by Germany-based MNEs is higher as the tax difference between Germany and their destination country gets wider.
Important destination countries of German FDI are Germany's neighboring ones.
Therefore, the empirical results of these studies are consistent with our prediction that the high-tax country attracts multinational production from the low-tax country when bilateral trade openness is high.\footnote{Strictly speaking, \cite{Overesch2009} and \cite{Goldbachetal2019} consider a different mechanism than ours in that they consider incremental investment by a representative MNE, while we consider discrete investment by a continuum of MNEs.
In the theoretical model of \cite{Overesch2009}, the amount of shifted profits is assumed to be related to that of capital invested by the affiliate.
Profit shifting from an affiliate in high-tax Germany to its parent in a low-tax source country reduces the required rate of return to the affiliate, leading to an increase in  investment by the affiliate.
\cite{Goldbachetal2019} apply the same argument to the case of Germany-based parents.}

\

\

\section{Extensions}

This section discusses four extensions of our basic model.

\subsection{Transfer-pricing regulation}

We established Propositions 1 and 2 under no transfer-pricing regulation, i.e., $\delta=0$.
A tighter regulation (higher $\delta$) limits the effectiveness of profit shifting such that it reduces the deviation of the transfer price from the true marginal cost.
Thus, MNEs find it less profitable to move to country 1, where there is a higher tax rate but little scope for transferring profits.
In fact, we can numerically confirm in Figure 3 that a higher $\delta$ makes less likely the concentration of production in the high-tax country 1.
When $\delta$ is sufficiently high such that $\delta = (t_1-t_2)/(1-t_1) = (0.3 -0.2)/(1-0.3)=0.143$ (dashed curve), there is no room for profit shifting ($g_i = \tau a$) so that the plant share in country 1 coincides with that in the case without transfer pricing (dashed curve in Figure 2) and never exceeds one-half.

\

\begin{center}
\includegraphics[scale=0.55]{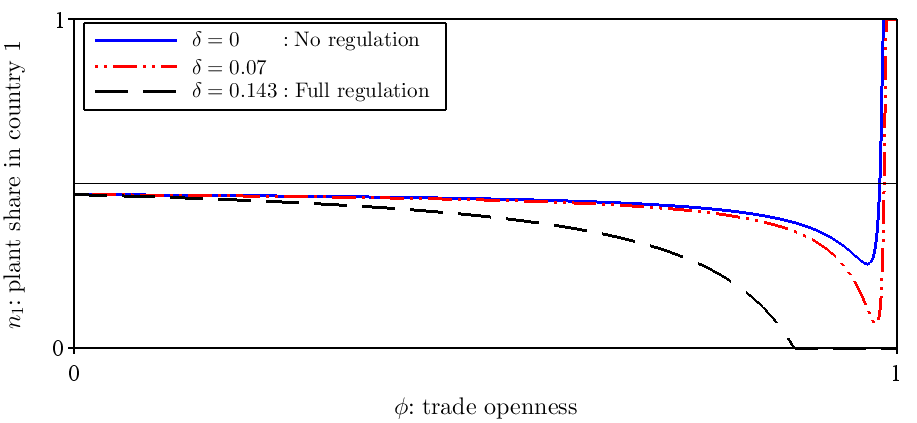} \\
Figure 3. \ Impact of transfer-pricing regulation \\
{\footnotesize {\it Notes:} \ Parameter values are $\sigma=5$; $t_1=0.3$; $t_2=0.2$; $s_1=0.5$; $\delta \in \{ 0, 0.07, 0.143\}$.}
\end{center}

\

\subsection{Pure exporters}

In the main analysis, we excluded the possibility of pure exporters, who export varieties without using distribution affiliates.
However, one may expect that it would be profitable for production plants in the low-tax country to serve the high-tax country by exporting directly to consumers, rather than via distribution affiliates.
By exporting while producing in the low-tax country, firms can save taxes and avoid setting a high selling price for the foreign market.
To examine this, we endogenize whether each firm becomes an MNE or a pure exporter.
The MNE uses two units of capital to establish a plant in one country and a distribution affiliate in the other, as in the basic model.
The pure exporter uses $1+\kappa$ units of capital to set up two plants in the same country, one for the home market and the other for the foreign market.
The benefit of becoming a pure exporter is modeled as low fixed costs, i.e., $\kappa \in [0, 1]$ and $1+\kappa \le 2$, following \cite{Helpmanetal2004}.

The timing of actions is modified as follows.
Firms first decide the location of a plant for the home market and then choose their organizational form, either an MNE or a pure exporter.
The choice of organizational form is equivalent to the choice of where to locate the second plant/affiliate.
The subsequent actions proceed as in the basic model.

This extended model is fully analyzed in Online Appendix F, where we confirm the aforementioned reasoning: firms locating the first plant in the low-tax country 2 always become pure exporters.
Given production in the low-tax country, it makes no sense to become an MNE because its distribution affiliates faces high taxes and is less competitive due to a higher transfer price of inputs.
By contrast, firms locating the first plant in the high-tax country 1 may become an MNE or a pure exporter.
The relative benefit of becoming an MNE instead of a pure exporter depends on the fixed cost of export $\kappa$ and trade openness.

When $\kappa$ is high, the benefit of becoming a pure exporter is small.
As trade openness rises, the high-tax country 1 first loses and then gains multinational production, as in the basic model.
By contrast, when $\kappa$ is low, firms in both countries become pure exporters and move away from the high-tax country 1, as openness rises.
As long as the benefit of becoming a pure exporter is not too large ($\kappa$ is not too small), we maintain the conclusion of full production agglomeration in the high-tax country for high openness.

\

\subsection{Asymmetric country size}

The result of the high-tax country 1 attracting more multinational production for high trade openness, in general, does not depend on the assumption of symmetric market size.
Our model inherits a common feature of NEG models that firms try to locate in large countries to save trade costs when exporting.
Thus, if country 1 is larger ($s_1>1/2$), the larger market size strengthens the incentive of MNEs to locate production there.
If country 1 is smaller ($s_1<1/2$), MNEs have less incentive to choose it for production.
Even in this case, however, the small high-tax country 1 may achieve full production agglomeration for high openness, where the trade-cost-saving motive is weak.
In Online Appendix G, we confirm that production plants are always agglomerated in the high-tax country 1 at $\phi=1$ except in the case where country 1 is extremely large.\footnote{In the special case where $s_1>\overline{s}_1 \equiv \sigma/(2\sigma-\Delta t_2)$, full production agglomeration in the large, high-tax country 1 does not occur.
This is because the transferable profits, which depend on the sales of distribution affiliates in country 2, are very small.
This exceptional case implies that the very small, low-tax country (roughly a tax-haven) may host a greater share of plants than its market-size share: $n_2 > s_2$.}

\subsection{Centralized decision making}

We assumed that MNEs have decentralized decision making, where foreign affiliates choose prices to maximize their own profits. 
Our main result holds true even if MNEs have centralized decision making, whereby they choose all prices to maximize global profits. 
Note that the direction of profit shifting does not change depending on the decision making style.
That is, foreign affiliates source goods from production plants by paying high (or low) transfer prices if they are in the low-tax country (or the high-tax country).
By locating in the low-tax country, foreign affiliates enjoy a higher price-cost margin than those located in the high-tax country ($p_{12}-g_1>p_{21}-g_2$) and earn larger profits.
As in the decentralized decision making case, 
profit shifting affects the profitability of foreign affiliates asymmetrically, leading to the agglomeration of production plants in the high-tax country.
See Online Appendix H for more details.

\

\section{Tax competition}

Here, we allow countries to choose their tax rate and compare the results of tax competition in the case with and without transfer pricing.
Throughout this section except for Section 5.3, we cast aside transfer-pricing regulation, i.e., $\delta=0$.
The objective function of government $i \in \{ 1, 2\}$ takes the form of
\begin{align*}
&G_i = TR_i - \frac{t_i}{\alpha_i(1-t_i)}, \\
\
&\text{where} \ \ TR_i \equiv t_i \cdot TB_i, \\
\
&\ \ \ \ \ \ \ \ \ \ TB_i \equiv N_i \pi_{ii} + N_j \pi_{ji}, \ \ \ \ i \neq j \in \{ 1, 2\}.
\end{align*}
The tax base for government $i$, $TB_i$, consists of the profits of home production plants, $N_i \pi_{ii}$, and those of foreign distribution affiliates, $N_j \pi_{ji}$.
The second term of $G_i$ is tax administration costs, where $\alpha_i>0$ captures its efficiency.
We assume that government 1 is more efficient in tax administration than government 2: $\alpha_1 > \alpha_2$.\footnote{Tax administration costs are well recognized as an important determinant of raising revenues (\citealp{OECD2017admin}; \citealp{ProfetaScabrosetti2017}).
\cite{OECD2017admin} states that ``Even small increases in compliance rates or compliance costs can have significant impacts on government revenues and the wider economy.'' (p.5)
In addition, this objective function captures the fundamental conflicts governments face: they attempt to raise tax revenues while maintaining a low tax rate, which is deemed a reduced-form objective that either a selfish or a benevolent government adopts (\citealp{BaldwinKrugman2004}).
See also \cite{BorckPfluger2006}; \cite{Hanetal2014}; and \cite{Kato2015} for similar specifications.
An alternative interpretation of the objective function is that the government is supported (or elected) by residents, who put a high value on public goods/services funded by tax revenues.
This specification is used in \cite{Krautheimetal2011} in the context of tax competition between a tax-haven and a non-tax-haven country.}
The two governments simultaneously and non-cooperatively decide their tax rates before the MNEs' location decisions.

\

\noindent
\subsection{No-transfer-pricing case} 
As a benchmark, we derive the equilibrium tax rates when transfer pricing is not allowed.
The inability to manipulate transfer prices implies that the transfer price must be equal to the true marginal cost: $g_i = \tau a$, in which case $\gamma_i=1$ holds.
This leads to zero profits from intra-firm trade: $(g_i -\tau a)q_{ij} = 0$ for $i \neq j \in \{ 1, 2\}$.
Combining these results with $\pi_{11}$ in Eq. (5-1) and $\pi_{21}$ in Eq. (5-2) yields tax revenues for government 1:
\begin{align*}
TR_1 &= t_1 TB_1 \\
\
&= t_1 [N_1 \pi_{11} + N_2 \pi_{21}] \\
\ 
&= t_1 \biggl[ N_1 \cdot \frac{\mu L_1}{\sigma(N_1 +\phi N_2)} 
+ N_2 \cdot \frac{\phi \mu L_1}{\sigma(N_1 + \phi N_2)}  \biggr] \\
\
&= t_1 \cdot \frac{ \mu L_1 (N_1 + \phi N_2) }{ \sigma(N_1 + \phi N_2) } \\
\
&= t_1 \cdot \frac{\mu L}{2\sigma},
\end{align*}
where $L_1=L_2=L/2$.
We can analogously derive tax revenues for government 2 as $TR_2 = t_2 TB_2 = t_2 \mu L/(2\sigma)$.
Notably, $TB_i$ depends on neither the share of plants ($n_i$), trade openness ($\phi$), nor taxes ($t_i$s).
Two factors explain that the tax base is independent of $n_i$, $\phi$ and $t_i$s.\footnote{The constant tax base does not result from the quasi-linear utility function with constant expenditure on manufacturing goods: $\sum_{i=1}^2 N_i p_{ij} q_{ij} = \mu L_j$.
In Online Appendix K, we confirm that the Cobb-Douglas utility function also yields a constant tax base.}
First, the total mass of plants and affiliates generating country 1's tax base is constant and is given by $N_1 + N_2 = n_1 L + (1-n_1)L = L$.
Second, the transferable profits of foreign affiliates depend on their sales in country 1, and thus, are limited by its residents' expenditure on manufacturing goods ($\sum_{i=1}^2 N_i p_{i1}q_{i1} = \mu L_1$).
The constant tax base then implies that governments do not benefit from the full agglomeration of plants.
Put differently, the concentration of multinational production in a country does not generate taxable agglomeration rents there, which sharply contrasts with the implication of the NEG models (\citealp{BaldwinKrugman2004}).

Government $i$ thus maximizes
\begin{align*}
G_i = \frac{\mu L t_i}{2\sigma} - \frac{t_i}{\alpha_i(1-t_i)}.
\end{align*}
Solving the first-order condition gives the equilibrium tax rate in country $i$:
\begin{align*}
&\widehat{t}_i = 1 - \sqrt{\frac{2 \sigma}{\alpha_i \mu L}}, \tag{7}
\end{align*}
where we use $\widehat{t}_i$ to represent the equilibrium value of $t_i$ in tax competition without transfer pricing. 
We can check the government payoff is positive: $G_i(t_i=\widehat{t}_i) = [\mu L/(2\sigma) -1/\alpha_i ]^2>0$.
The equilibrium tax rate increases with $\alpha_i$, reflecting the fact that a government with efficient tax administration can easily raise taxes.
In fact, government 1 sets a higher tax rate than government 2, i.e., $\widehat{t}_1 > \widehat{t}_2$.
Because the government objective functions depend on neither the plant share ($n_i$) nor trade openness ($\phi$), the equilibrium tax rates given by Eq. (7) are unique for any $n_i$ and $\phi$.
An excessively high (or low) $\alpha_i$ makes the equilibrium tax rate too high (or too low).
To ensure $\widehat{t}_i \in (0, 1/2)$ that guarantees positive total pre-tax profits (see Section 2.1), we assume $\alpha_i \in (2\sigma/(\mu L), 3\sigma/(\mu L))$.

Without transfer pricing, a higher tax rate of a country simply discourages production plants from locating there.
This tendency is more pronounced when trade openness is high.
In particular, all plants move away from the high-tax country 1 to the low-tax country 2, when the level of openness is higher than the agglomeration threshold $\widehat{\phi}^S$:
\begin{align*}
\widehat{\phi}^S &= \frac{1-\widehat{t}_1}{1-\widehat{t}_2} \in (0, 1), \tag{8}
\end{align*}
The situation of full production agglomeration in country 2, i.e., $n_1 = 0$, is shown in Figure 4 (dashed line).
At $\phi=1$, MNEs are indifferent to the production location because, unlike the transfer-pricing case, the pre-tax profits from the home and foreign markets are the same when there are no trade costs.

These results are summarized as follows (see Online Appendix I for the proof).

\pagebreak

\noindent
{\bf Proposition 3 (tax competition without transfer pricing).} {\it Assume that two countries are of equal size ($s_1=1/2$) and that government 1 has a more efficient tax administration than government 2: $\alpha_1  > \alpha_2$, where $\alpha_i \in (2\sigma/(\mu L), 3\sigma/(\mu L))$.
As a result of tax competition without transfer pricing, government 1 always sets a higher tax rate and hosts a smaller share of plants than government 2, i.e., $\widehat{t}_1 > \widehat{t}_2$, given by Eq. (7).
There exists a threshold value of trade openness, $\widehat{\phi}^S$, such that the following holds:
\begin{itemize}
\item[(i).] If $\phi \in [0, \widehat{\phi}^S)$, the high-tax country 1 hosts a smaller share of plants than the low-tax country 2, i.e., $n_1 \in (0, 1/2)$.
\item[(ii).] If $\phi \in [\widehat{\phi}^S, 1)$, the high-tax country 1 loses all plants, i.e., $n_1 =0$.
\item[(iii).] If $\phi =1$, any plant distribution can be the long-run equilibrium, i.e., $n_1 \in [0, 1]$.
\end{itemize}}

\

\begin{center}
\includegraphics[scale=1.0]{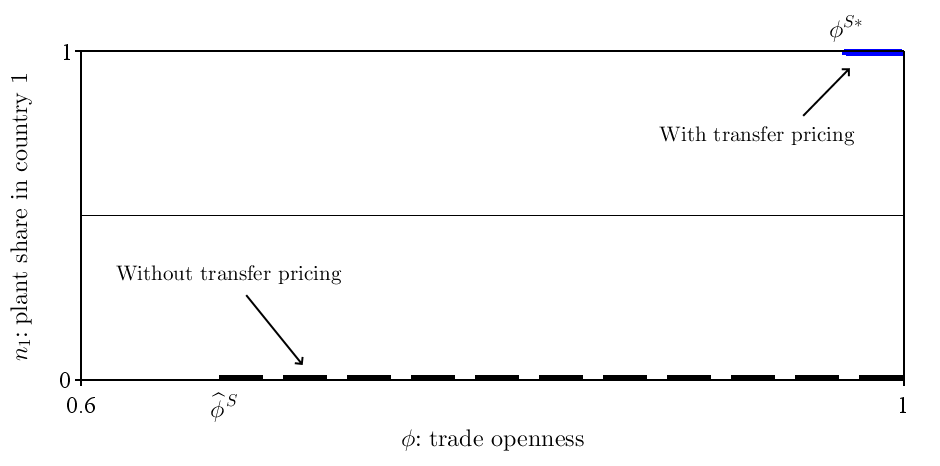} \\
Figure 4. \ Plant share under tax competition \\
{\footnotesize {\it Notes}: \ Parameter values are $\sigma=5$; $t_1=0.3$; $t_2=0.2$; $\delta=0$; $s_1=0.5$.}
\end{center}

\

\noindent
\subsection{Transfer-pricing case}
We examine how these results change if MNEs can use transfer pricing.
The analysis in Section 3 tells that when trade openness is high enough such that $\phi \in [\phi^S, 1]$, all production plants are agglomerated in country 1, as long as its tax rate is higher than that of country 2 ($t_1>t_2$).
This situation is illustrated in Figure 4 (thick line).\footnote{We obtain the agglomeration threshold $\phi^{S*}$ in Figure 4 by evaluating $\phi^{S}$ at the equilibrium tax rate $\{ t_i^*\}_{i=1}^2$, which will be discussed shortly.}
Here, we focus on the range $\phi \in [\phi^S, 1]$ and derive conditions under which full production agglomeration emerges as a result of tax competition.

Supposing that all plants are in country 1 ($N_1=n_1 L = L$), we can derive tax revenues for government 1 as
\begin{align*}
TR_1 &= t_1 TB_1 \\
\ 
&= t_1 \left[ \frac{\mu L}{2\sigma} + N_1 \frac{(\sigma-1)\Delta{t}_1}{\sigma} \frac{\phi \gamma_1 \mu L}{2\sigma(\phi \gamma_1 N_1 +N_2)} \right] \\
\
&= t_1 \Bigg[ \frac{\mu L}{2\sigma} + \underbrace{ \frac{(\sigma-1)\Delta{t}_1}{\sigma} \frac{ \mu L }{2\sigma} }_{\text{Shifted profits of home plants} \ < 0} \Bigg],
\end{align*}
where $L_1=L_2=L/2$; $N_1= n_1 L = L$; and $N_2 = (1-n_1)L = 0$.
By contrast, tax revenues for government 2 remain unchanged because there are no production plants there, which transfer profits.
We obtain
\begin{align*}
TR_2 &= t_2 TB_2 \\
\
&= t_2 \left[ \frac{\mu L}{2\sigma} +N_2 \frac{(\sigma-1)\Delta{t}_2 }{\sigma} \frac{\phi \gamma_2 \mu L}{2\sigma(N_1 +\phi \gamma_2 N_2)} \right] \\
\
&= t_2 \cdot \frac{\mu L}{2\sigma}.
\end{align*}

The governments' payoffs then become
\begin{align*}
&G_1 = \frac{\mu L t_1}{2\sigma}\bigg[ 1 + \frac{(\sigma-1)\Delta{t}_1}{\sigma} \bigg] - \frac{t_1}{\alpha_1(1-t_1)}, \\
\
&G_2 = \frac{\mu L t_2}{2\sigma} - \frac{t_2}{\alpha_2(1-t_2)},
\end{align*}
where $G_1$ now involves the tax difference due to the presence of shifted profits of home plants.
To prevent this tax base erosion, government 1 has a stronger incentive to lower its tax rate than it does in the no-transfer-pricing case. 
By contrast, as country 2 does not have any plants that receive shifted profits, $G_2$ is the same as in the no-transfer-pricing case.\footnote{Profits that foreign affiliates transfer from/to plants are indeed included in the first term of $TB_i$ and do not show up explicitly. See Online Appendix E for more details.}

From the first-order conditions, we obtain
\begin{align*}
&t_1^* = 1 - \sqrt{\frac{2\sigma}{\alpha_1 \mu L} + \frac{(\sigma-1)(\sqrt{2 \sigma \mu L/\alpha_2} -2 \sigma/\alpha_1)}{\mu L (2\sigma -1) }}, \tag{9-1} \\
\
&t_2^* = 1 - \sqrt{\frac{2 \sigma}{\alpha_2 \mu L}} \ \ \ \ (= \widehat{t}_2), \tag{9-2}
\end{align*}
where we use $t_i^*$ to represent the equilibrium value of $t_i$ in tax competition with transfer pricing. 
We can confirm that $t_1^*$ and $t_2^*$ are in $(0, 1/2)$ under our assumption that $\alpha_1 >  \alpha_2$, where $\alpha_i \in (2\sigma/(\mu L), 3\sigma/(\mu L))$.
To be consistent with the full agglomeration of plants in country 1, $t_1^*$ must be higher than $t_2^*$.
This requires a sufficiently high $\alpha_1$ such that $\alpha_1 \in (\alpha^*, 3\sigma/(\mu L))$ with $\alpha^* > \alpha_2$.
We can also check that $t_1^*$ is lower than the equilibrium tax rate without transfer pricing $\widehat{t}_1$.
Introducing profit shifting intensifies tax competition in the sense that the tax difference becomes narrower.\footnote{If instead $\alpha_1$ is higher than $\alpha_2$ but is low enough such that $\alpha_1 \in (\alpha_2, \alpha^*]$, $t_1^*$ is equal to or lower than $t_2^*$ and full production agglomeration in country 1 does not occur.
In this case, government 1 chooses to mimic government 2's tax rate ($t_1^* = t_2^*$), whereas government 2 chooses the same tax rate as it does in the no-transfer-pricing case ($t_2^* = \widehat{t}_2$).
This leads to a symmetric distribution of plants ($n_1 = 1/2$).}

Thus, our main result that profit shifting leads to production agglomeration in the high-tax country for high trade openness (Proposition 1) carries over to a tax-competition framework when countries are quite different in their efficiency in tax administration.
We can formally prove the following proposition.

\

\noindent
{\bf Proposition 4 (Tax competition with transfer pricing).} {\it Assume that $s_1=1/2$, $\alpha_1 > \alpha_2$, where $\alpha_i \in (2\sigma/(\mu L), 3\sigma/(\mu L))$, as in Proposition 3.
Consider tax competition with transfer pricing under sufficiently high trade openness such that $\phi \in [\phi^{S}, 1]$, where $\phi^{S}$ is the agglomeration threshold.
Then, there exists a unique pair of equilibrium tax rates, $\{ t_i^* \}_{i=1}^2$ given by Eqs. (9-1) and (9-2), which satisfy the following:}
\begin{itemize}
\item[{\it (i).}] ({\bf The tax rates of country 1 versus country 2}) \ \ {\it Country 1's tax rate is higher than country 2's $(t_1^*>t_2^*$) if the tax-administration efficiency of government 1 is sufficiently high: $\alpha_1 > \alpha^*$. 
All production plants are located in country 1 ($n_1 = 1$) for $\phi > \phi^{S*}$, where $\phi^{S*}$ is the agglomeration threshold evaluated at the equilibrium tax rates.}
\item[{\it (ii).}] ({\bf Tax rates with and without transfer pricing}) \ \ {\it Assume $\alpha_1 > \alpha^*$ and $\phi > \phi^{S*}$.
Compared to the no-transfer-pricing case, country 1's tax rate and payoff decrease ($t_1^*<\widehat{t}_1$; $G_1(t_1^*) < G_1(\widehat{t}_1)$), whereas country 2's tax rate and payoff are unchanged ($t_2^*=\widehat{t}_2$;  $G_2(t_2^*) = G_2(\widehat{t}_2)$).
That is, introducing transfer pricing makes tax competition tougher ($0 < t_1^* -t_2^* < \widehat{t}_1 - \widehat{t}_2$) and leaves the world worse off ($G_1(t_1^*) + G_2(t_2^*) < G_1(\widehat{t}_1) + G_2(\widehat{t}_2)$).} 
\end{itemize}

\

\noindent See Online Appendix J for the proof.
We note that, unlike the no-transfer-pricing case, full production agglomeration occurs even with completely free trade ($\phi=1$).
This is because MNEs can differentiate the two selling prices using transfer prices dependent on the international tax difference.
In Online Appendix K, we also confirm that when the utility function takes the Cobb-Douglas form, the results remain qualitatively similar to Proposition 4.

Our conclusion that profit shifting pushes taxes downward (Proposition 4(ii)) is consistent with \cite{HauflerSchjelderup2000}, who employ a framework of perfect competition with symmetric countries.\footnote{\cite{AgrawalWildasin2020} also show that globalization, defined by a decline in relocation costs, leads to tougher tax competition
in a linear spatial model where agglomeration is exogenously given.}
By contrast, \cite{Stowhase2005, Stowhase2013} obtain the opposite result: introducing profit shifting softens tax competition by increasing the equilibrium tax rates of countries.\footnote{\cite{BeckerRiedel2013} obtain a similar result, although MNEs in their model cannot shift profits for tax-saving purposes.}
In the presence of profit shifting, governments chase the shifted profits and intensify tax competition.
Meanwhile, MNEs can save tax payments regardless of their locations and become less sensitive to international tax differences, thereby making tax competition less severe.
In \cite{Stowhase2005, Stowhase2013}, the latter effect dominates the former, 
whereas the opposite is true in our model.
These differing results may come from the fact that while \cite{Stowhase2005, Stowhase2013} consider a representative firm or a monopoly firm, we consider a continuum of MNEs competing with each other.
The competitive environment strengthens the tax-saving incentive, and thus, increases the tax base sensitivity in the high-tax country hosting full production agglomeration.

Our result that the high-tax country achieves full production agglomeration (Proposition 4(i)) resembles the one in \cite{BaldwinKrugman2004} and other NEG models; however, their mechanism crucially differs from ours.
In \cite{BaldwinKrugman2004},
production agglomeration generates taxable rents, which are non-monotonic in terms of trade openness $\phi$.
A higher $\phi$ may expand the rents and soften tax competition.
By contrast, in our model with profit shifting, production agglomeration is harmful to a country by inducing the erosion of its tax base.
One government sets a higher tax rate than the other, not because it wants to tax agglomeration rents, but because it has a more efficient tax administration.

\

\subsection{Coordination of transfer-pricing regulation}

Based on Proposition 4(ii) that introducing transfer pricing leaves the world worse off, one may think that international coordination of transfer-pricing regulation would make the world better off.
We show that this is indeed possible as a result of mutual agreement between the two governments.
Note that international coordination of {\it tax rates} is difficult because both governments set the dominant-strategy equilibrium tax rate, and thus, do not have any incentive to change it.

Adding transfer-pricing regulation $\delta$ modifies the governments' payoffs as
\begin{align*}
&G_1 = \frac{\mu L t_1}{2\sigma}\bigg[ 1 + \frac{(\sigma-1)(\Delta{t}_1 +\delta)}{\sigma} \bigg] - \frac{ t_1}{\alpha_1(1-t_1)}, \\
\
&G_2 = \frac{\mu L t_2}{2\sigma} - \frac{t_2}{\alpha_2(1-t_2)},
\end{align*}
where government 2's payoff, $G_2$, is unchanged since there is no plant in country 2.
The associated equilibrium tax rates are
\begin{align*}
&t_1^{**} = 1 - \sqrt{\frac{2 \sigma}{\alpha_1 \mu L} + \frac{(\sigma-1)[ \sqrt{2 \sigma \mu L/\alpha_2} -2 \sigma(1+\delta)/\alpha_1] }{\mu L [2\sigma -1 +\delta(\sigma-1)]}}, \\
\
&t_2^{**} = 1 - \sqrt{\frac{2\sigma}{\alpha_2 \mu L}} \ \ \ \ (= t_2^* = \widehat{t}_2).
\end{align*}
A tighter regulation (higher $\delta$) increases $t_1^{**}$, and thus, raises government 1's payoff.
Since $\delta$ does not enter $G_2$, government 2 is indifferent to the stringency of regulation.\footnote{Here, we do not relate the stringency of regulation $\delta$ and the efficiency level of tax administration $\alpha_i$.
It is reasonable to think that $\delta$ increases with $\alpha_i$ because a more efficient government prevents its tax base erosion more effectively.
In Online Appendix L, we check that $t_1^{**}$ rises as the correlation of $\delta$ with $\alpha_i$ increases.}
These observations suggest the possibility of a Pareto improvement through the coordination of transfer-pricing regulation.
Both governments agree on tightening the regulation to make transfer pricing impossible, i.e., $\Delta t_1 + \delta = 0$ or $\delta = (t_1-t_2)/(1-t_2)$.
This result is summarized as follows.

\

\noindent
{\bf Proposition 5 (Transfer-pricing regulation).} {\it Consider tax competition with transfer pricing, as described in Proposition 4.
International coordination of regulation that prohibits transfer pricing, i.e., $\delta = (t_1-t_2)/(1-t_2)$, makes the world better off and is possible based on mutual agreement between the two countries.}

\

Studies on competition for profit-shifting multinationals typically find that setting transfer-pricing regulation leads to a prisoners' dilemma situation (e.g., \citealp{Peraltaetal2006}; \citealp{Hauck2019}; \citealp{HindriksNishimura2021}).\footnote{\cite{HindriksNishimura2021} point out two key drivers that help  governments agree on a cooperated outcome: (i) the complementarity of countries' efforts for regulation and (ii) tax leadership.}
Namely, governments never agree on international coordination for maximizing world payoffs because unilateral deviation from the agreed level of regulation always improves the national payoff.
In our model, however, government 2 is indifferent to the level of regulation, and thus, has no objection because it has no taxable profits that production plants try to shift to country 1.
When trade openness is sufficiently high such that plants fully agglomerate in a country with efficient tax administration, international cooperation on regulation is possible and sustainable. 

\

\section{Conclusion}

Countries with lower corporate tax rates are expected to host more multinational production.
However, this simplistic view may be challenged because economic integration, marked by falling trade costs (or rising trade openness), allows for profit shifting, and thus, may change the location incentives of multinationals.
To investigate this, we introduced transfer pricing into a simple two-country model of trade and geography.

With low trade openness, a low-tax country attracts more production plants than a high-tax country.
With high trade openness, this pattern completely reverses and production agglomerates in the high-tax country.
When high openness helps intra-firm trade expand, MNEs can use transfer pricing as both a strategic and a profit-shifting device.
To shift profits, the transfer price from the high-tax to the low-tax country is set low, whereas that from the low-tax to the high-tax country is set high.
This transfer-pricing strategy lowers the input cost of distribution affiliates in the low-tax country, and thus, makes them competitive.
By contrast, distribution affiliates in the high-tax country become less competitive due to the high input cost.
Therefore, MNEs prefer to locate production in the high-tax country and distribution in the low-tax country.

The main results carry over to a tax-competition framework where countries non-cooperatively choose their tax rates.
The difference in equilibrium tax rates between countries results from the difference in the level of efficiency in tax administration.
Unlike standard NEG models, the country hosting all production plants does not enjoy agglomeration rents from it; rather, it faces tax base erosion due to profit shifting.
Another finding is that introducing profit shifting makes tax competition fiercer by reducing the high-tax country's equilibrium tax rate.

We test our prediction using bilateral FDI data from 23 OECD countries from 2008 to 2016.
The empirical exercise supports the non-monotonic effect of economic integration on the share of production affiliates out of affiliates in all sectors. 
Furthermore, supporting evidence can also be found in \cite{Overesch2009} and \cite{Goldbachetal2019}.
For example, \cite{Overesch2009} confirms that foreign affiliates in Germany invest more, as the cross-country tax difference between their source countries and high-tax Germany increases.

Although our model is admittedly stylized, we believe that it is versatile enough to accommodate further extensions.
An interesting extension is to consider various channels of profit shifting such as licensing fees for intellectual property rights and the choice between equity and debt financing.
While we solely focus on the transfer pricing of tangible goods, 
these alternative channels are equally important in reality.
Adding these channels into our model may yield different implications for production location and tax revenues.
Another extension is to examine the impact of different international tax systems, such as separate accounting and formula apportionment.
Which tax system prevents profit shifting more effectively may differ depending on the degree of economic integration.
We leave these avenues for future research.

\


\pagebreak

\begin{spacing}{1}

\section*{Appendices}

\begingroup
\subsection*{Appendix 1. \ List of key variables} 
\setlength{\tabcolsep}{10pt} 
\renewcommand{\arraystretch}{1.15} 
\begin{table}[H] 
\footnotesize
	\begin{tabular}{cl}
		\cline{1-2}
		Variable & Description \\ \cline{1-2} \hline \hline
		$L_i$ & Population in country $i$   \\ \cline{1-2}
		$K_i$ & Capital endowment in country $i$   \\ \cline{1-2}
		$\kappa$ & Capital that pure exporters need to supply to a foreign market (Section 4.2) \\ \cline{1-2}
		$s_i$ & Share of population and capital endowment in country $i$   \\ \cline{1-2}
		$w_i$ & Wage in country $i$  \\ \cline{1-2}
		$a$ & Unit labor requirement for the differentiated varieties \\ \cline{1-2}
		$R_i$ & Reward to capital invested in MNEs with production in country $i$   \\ \cline{1-2}
		$\mu$ & Intensity of the preference for the differentiated goods  \\ \cline{1-2}
		$\sigma$ & Elasticity of substitution over differentiated varieties   \\ \cline{1-2}
		$\widetilde{q}_{ij}(\omega)$ & Individual demand for the variety $\omega$ produced in country $i$ and consumed in country $j$ \\ \cline{1-2}
		$q_{ij}(\omega)$ & Aggregate demand for the variety $\omega$ produced in country $i$ and consumed in country $j$ \\ \cline{1-2}
		$Q_i$ & CES aggregator of the differentiated varieties in country $i$   \\ \cline{1-2}
		$p_{ij}(\omega)$ & Price of the variety $\omega$ produced in country $i$ and consumed in country $j$   \\ \cline{1-2}
		$P_i$ & Price index of the differentiated varieties in country $i$   \\ \cline{1-2}
		$n_i$ & Share of production plants in country $i$   \\ \cline{1-2}
		$N_i$ & Mass of MNEs with production in country $i$ $(=n_i L)$  \\ \cline{1-2}
		$g_i$ & Transfer price set by MNEs headquartered in country $i$   \\ \cline{1-2}
		$\delta$ & Stringency of transfer pricing regulation   \\ \cline{1-2}
		$\pi_{ii}$ & Pre-tax operating profits of a production plant of MNEs with production in country $i$   \\ \cline{1-2}
		$\pi_{ij}$ & Pre-tax operating profits of a distribution affiliate of MNEs with production in country $i$   \\ \cline{1-2}
		$\Pi_i$ & Global post-tax profits of MNEs with production in country $i$  \\ \cline{1-2}
		$t_i$ & Corporate tax rate in country $i$  \\ \cline{1-2}
		$\Delta t_i$ & Weighted tax gap $\left( \equiv (t_j-t_i)/(1-t_i) \right)$ \\ \cline{1-2}
		$\widehat{t}_i$ & Equilibrium tax rate in country $i$ without transfer pricing (Section 5.1)   \\ \cline{1-2}
		$t_i^*$ & Equilibrium tax rate in country $i$ under tax competition with transfer pricing (Section 5.2)   \\ \cline{1-2}
		$t_i^{**}$ & Equilibrium tax rate in country $i$ under tax competition with transfer price regulation (Section 5.3)   \\ \cline{1-2}
		$\gamma_i$  & Bundle of parameters $\left( \gamma_1 \equiv \left( \dfrac{\sigma(1+\delta)}{\sigma-\Delta t_1+\delta(\sigma-1)} \right)^{1-\sigma}, \ \ \ \ \gamma_2 \equiv \left( \dfrac{\sigma(1-\delta)}{\sigma-\Delta t_2-\delta(\sigma-1)} \right)^{1-\sigma}  \right)$ \\ \cline{1-2}
		$\tau$ & Iceberg trade cost of the differentiated goods   \\ \cline{1-2}
		$\phi$ & Trade openness $\left(\equiv \tau^{1-\sigma}\right)$\\ \cline{1-2}
		$\phi^S$ & Agglomeration threshold above which level of $\phi$ all plants are located in one country ($n_1 \in \{ 0, 1\}$) \\ \cline{1-2}
		$\phi^\dag$ & Threshold of $\phi$ such that $n_1=1/2$ holds  \\ \cline{1-2}
		$\phi^\#$ & Threshold of $\phi$ such that $d n_1/d \phi=0$ holds \\ \cline{1-2}
		$\widehat{\phi}^S$ & Agglomeration threshold under tax competition without transfer pricing (Section 5.2) \\ \cline{1-2}		
		$\phi^{S*}$ & Agglomeration threshold under tax competition with transfer pricing (Section 5.2) \\ \cline{1-2}		
		$TB_i$ & Tax base in country $i$  \\ \cline{1-2}
		$TR_i$ & Tax revenue in country $i$ \\ \cline{1-2}
		$G_i$ & Objective function of the government in country $i$   \\ \cline{1-2}
		$\alpha_i$ & Efficiency measure of tax administration \\ \cline{1-2}	 
	\end{tabular} 
\end{table}
\endgroup

\end{spacing}

\begin{spacing}{1.0}
\subsection*{Appendix 2. \ Data}

Sources for the variables used in the paper and their descriptions are as follows.

\

{\it Foreign affiliates.} \ We take bilateral outward ``Number of enterprises'' data from Outward activity of multinationals by country of location--ISIC Rev 4. in OECD Statistics, covering the period from 2007 to 2017.
Although the data is available by sector, we only use data on ``TOTAL BUSINESS SECTOR'' and ``MANUFACTURING'' because many of sector-level data are missing.
The sample consists of 23 OECD counties: Australia, Austria, Belgium, Canada, Czech Republic, Germany, Finland, France, Greece, Hungary, Israel, Italy, Japan, South Korea, Norway, Poland, Portugal, Slovakia, Slovenia, Spain, Sweden, United Kingdom, United States.
We exclude observations with missing values and/or zero affiliates from the sample, so that the sample period is reduced to the period 2008 to 2016.
Note that four European countries identified as tax havens by \cite{Zucman2014}, i.e., Ireland, Luxembourg, the Netherlands, and Switzerland, are not included in the sample.
In regression analysis, we define a host-source-year level ``Manufacturing affiliate share'' as the share of affiliates in ``MANUFACTURING'' out of those in ``TOTAL BUSINESS SECTOR.''

\

{\it Corporate tax rates.} \ Data on statutory corporate tax rates and effective average tax rates are from Centre for Business Taxation Tax Database 2017 (\citealp{Habu2017}).
In regression analysis, we convert them into a percentage value and take their difference between host country $h$ and source country $s$, $\Delta TAX_{h,s,t} = TAX_{h,t}-TAX_{s,t}$, where with an abuse of notation we here use $t$ for time subscript.

\

{\it Trade openness.} \ We take the bilateral distance measure from \cite{MayerZignago2011} and define the trade openness as the inverse of the log of the bilateral geodesic distance between the main city of the two countries.
As an alternative measure, we use the inverse of the log of the bilateral trade costs in manufacturing sector constructed by the ESCAP-World Bank Trade Cost Database (\citealp{Arvisetal2016}).
This alternative measure is time variant and computed using data on bilateral trade flows {\`a} la \cite{HeadRies2001}; and \cite{Novy2013}.

\

{\it Pro-business index.} \ As a measure of pro-business conditions, we use the ``Index of Economic Freedom'' published by the Heritage Foundation.
The Index includes 12 sub-indices, each of which ranges from zero to one and captures the rule of law, government size, regulation efficiency, and market openness. 
Among the 12 sub-indices, we exclude the ``tax burden'' index and redefine a simple average of the other 11 sub-indices as the pro-business index.
In regression analysis, we use the difference of this index between a host and a source country, $\Delta (\text{Pro business})_{h,s,t}$, as the IV for their corporate tax difference.

\

{\it Accountability index.} \ As a measure of political process, we use a summary index of government accountability constructed by the World Governance Indicators (WGI) of the World Bank.
This accountability index is a mean of 21 policy indices reported in the Institutional Profiles Database of the CEPII.
The sub-indices measure e.g., freedom of elections at national level, reliability of state budget, and freedom of association, capturing perceptions of the extent to which a country's citizens are able to participate in selecting their government, as well as freedom of expression, freedom of association, and a free media.\footnote{The 21 sub-indices are listed under name of ``IPD'' in \url{https://info.worldbank.org/governance/wgi/Home/downLoadFile?fileName=va.pdf}, accessed on 10 August 2022.
Both the accountability index (e.g., ``IPD060708VA'') and the sub-indices are given in \url{https://info.worldbank.org/governance/wgi/Home/downLoadFile?fileName=IPD.xlsx}, accessed on 10 August 2022.}
The original index constructed by the WGI ranges from $0$ to $1$, which we transform from $0$ to $100$.
We use the difference of this index between a host and a source country, $\Delta (\text{Accountability})_{h,s,t}$, for the IV of their corporate tax difference.

\

{\it Other variables.} \ We respectively use the growth rate ($\%$) of unit labor costs hours based as a measure of labor costs and the growth rate ($\%$) of GDP per hour worked as a measure of productivity.
In regression analysis, we take the difference of these values between a host and a source country: $\Delta (\text{Labor costs})_{h,s,t}$ and $\Delta (\text{Productivity})_{h,s,t}$.
These data and real GDP used to draw Figure 1 are from OECD statistics.

\end{spacing}

Summary statistics for the variables used in regression is provided in Table A1 in Appendix.

\begingroup
\begin{center}
Table A1 \\
Summary statistics \\
\begin{tabular}{@{\extracolsep{5pt}}lccccc} 
\\[-1.8ex]\hline \hline  \\[-1.8ex] 
 & \multicolumn{1}{c}{N} & \multicolumn{1}{c}{Mean} & \multicolumn{1}{c}{St. Dev.} & \multicolumn{1}{c}{Min} & \multicolumn{1}{c}{Max} \\ 
\hline \\[-1.8ex] 
Manufacturing affiliate share & 1,833 & 0.345 & 0.167 & 0.022 & 1.000 \\ 
$\Delta TAX_{h,s,t}$: statutory rate \ \ ($\%$point) & 1,833 & $-$0.654 & 9.043 & $-$21.760 & 23.460 \\ 
$\Delta TAX_{h,s,t}$: effective average rate \ \ ($\%$point) & 1,833 & $-$0.667 & 7.926 & $-$20.191 & 20.028 \\ 
$\phi_{h,s}$: 1/$\log$(distance) & 1,833 & 0.135 & 0.021 & 0.102 & 0.245 \\ 
$\phi_{h,s,t}$: 1/$\log$(trade costs) & 1,630 & 0.238 & 0.046 & 0.183 & 1.074 \\
$\Delta (\text{Labor cost})_{h,s,t}$ \ \ ($\%$point) & 1,833 & 0.330 & 6.729 & $-$31.025 & 31.454 \\ 
$\Delta (\text{Productivity})_{h,s,t}$ \ \ ($\%$point) & 1,833 & 0.264 & 1.975 & $-$9.554 & 11.407 \\ 
$\Delta (\text{Pro business})_{h,s,t}$ \ \ ($\%$point) & 1,833 & 0.392 & 8.398 & $-$25.264 & 25.745 \\ 
$\Delta (\text{Accountability})_{h,s,t}$ & 1,825 & $-$0.906 & 7.098 & $-$21.825 & 21.429 \\ 
\hline 
\hline \\[-4ex] 
\end{tabular} 
\end{center}
\endgroup

\pagebreak

\begin{spacing}{1.0}
\subsection*{Appendix 2. \ Robustness check for the non-monotonic effect}

In the regression analysis shown in Table 1, we use the geodesic distance to construct the measure of bilateral trade openness $\phi_{h,s}$.
As an alternative measure, we use the inverse of the log of trade costs reported by the ESCAP-World Bank Trade Cost Database (\citealp{Arvisetal2016}).
This alternative measure is time variant.
We also include the interaction terms of the control variables with trade openness and its squared, in addition to the level terms of trade openness and its squared.
The regression results are provided in Table A2.
As a further robustness check, we use a one-year lagged tax difference to mitigate the endogeneity concern on taxes and FDI flows.
The regression results with lagged tax difference are in Table A3.
In all specifications, the coefficient of $\Delta TAX_{h,s,t} \cdot \phi_{h,s,t}$ is negative and that of $\Delta TAX_{h,s,t} \cdot \phi_{h,s,t}^2$ is positive with strong statistical significance, confirming the non-monotonic effect of economic integration on multinational production.
\end{spacing}

\pagebreak

\begingroup
\setlength{\tabcolsep}{6pt} 
\renewcommand{\arraystretch}{0.8} 
\begin{center}
Table A2 \\
Robustness check for the non-monotonic effect \\
\vspace{0.3cm} \small
\begin{tabular}{lccc}
\hline \hline
	 & \multicolumn{1}{c}{Statutory tax rate} & & \multicolumn{1}{c}{Effective average tax rate} \\ \cline{2-2} \cline{4-4}
	 	 	 & (1) & &  (2) \\ \hline
  $\Delta TAX_{h ,s, t} \cdot \phi_{h, s, t}$ & $-$0.122*** & &   $-$0.147*** \\ 
  & (0.047) & &  (0.051) \\ 
  $\Delta TAX_{h ,s, t} \cdot \phi_{h, s, t}^2$ & 0.150** & &   0.184*** \\ 
  & (0.059) & &  (0.067) \\ [0.4cm]
  $\Delta TAX_{h ,s, t}$ & 0.014 & &  0.018 \\ 
  & (0.121) & &  (0.112) \\ 
  $\phi_{h,s,t}$ & $-$5.515*** & &  $-$5.546*** \\ 
  & (0.567) & &  (0.573) \\ 
  $\phi_{h,s,t}^2$ & 6.091*** & &  6.146*** \\ 
  & (0.788) & &  (0.802) \\ 
  $\Delta (\text{Labor cost})_{h ,s, t}$ & $-$0.051*** & &  $-$0.050** \\ 
  & (0.018) & &  (0.022) \\ 
  $\Delta (\text{Labor cost})_{h ,s, t} \cdot \phi_{h,s,t}$ & 0.405*** & &  0.402*** \\ 
  & (0.104) & &  (0.104) \\ 
  $\Delta (\text{Labor cost})_{h ,s, t} \cdot \phi_{h,s,t}^2$ & $-$0.780*** & &  $-$0.776*** \\ 
  & (0.166) & &  (0.166) \\  
  $\Delta (\text{Productivity})_{h ,s, t}$ & 0.195*** & &  0.193*** \\ 
  & (0.061) & &  (0.061) \\ 
  $\Delta (\text{Productivity})_{h,s, t} \cdot \phi_{h,s,t}$  & $-$1.296*** & &  $-$1.283*** \\ 
  & (0.381) & &  (0.382) \\ 
  $\Delta (\text{Productivity})_{h,s, t} \cdot \phi_{h,s,t}^2$  & 2.157*** & &  2.142*** \\ 
  & (0.549) & &  (0.550) \\ 
  $\Delta (\text{Pro business})_{h ,s, t}$ & $0.032$ & &  $0.029$ \\ 
  & (0.047) & &  (0.042) \\ 
  $\Delta (\text{Pro business})_{h ,s, t} \cdot \phi_{h,s,t}$ & $-$0.304*** & &  $-$0.286*** \\ 
  & (0.094) & &  (0.089) \\ 
  $\Delta (\text{Pro business})_{h ,s, t} \cdot \phi_{h,s,t}^2$  & 0.618*** & &  0.591*** \\ 
  & (0.155) & &  (0.147) \\ \hline
	Host country--year dummy & $\checkmark$ & &  $\checkmark$ \\
	Source country--year dummy & $\checkmark$ & &  $\checkmark$ \\ 
	Year dummy & $\checkmark$ & &  $\checkmark$  \\ 
	Observations & 1,630 & &  1,630 \\ 
	$R^2$ & 0.597 & &  0.597 \\ \hline \hline 
\end{tabular} \\ 
\begin{flushleft} {\footnotesize
\textit{Notes}: The dependent variable is the share of a source country's manufacturing affiliates in a host country out of the source's affiliates in all sectors in the host in a year.
Standard errors clustered at the host country-year level are in parentheses.
Unlike Table 1, we use the inverse of the log of the Head-Ries index as a measure of trade openness $\phi_{h,s,t}$. \\
***Significant at the 1$\%$ level; **Significant at the 5$\%$ level; *Significant at the 10$\%$ level.}
\end{flushleft}
\end{center}
\endgroup

\pagebreak

\begingroup
\setlength{\tabcolsep}{6pt} 
\renewcommand{\arraystretch}{0.8} 
\begin{center}
Table A3 \\
Robustness check for the non-monotonic effect: lagged tax difference \\
\vspace{0.3cm} \small
\begin{tabular}{lccc}
\hline \hline
	 & \multicolumn{1}{c}{Statutory tax rate} & & \multicolumn{1}{c}{Effective average tax rate} \\ \cline{2-2} \cline{4-4}
	 	 	 & (1) & &  (2) \\ \hline
  $\Delta TAX_{h ,s, t-1} \cdot \phi_{h, s, t}$ & $-$0.122*** & &   $-$0.147*** \\ 
  & (0.047) & &  (0.051) \\ 
  $\Delta TAX_{h ,s, t-1} \cdot \phi_{h, s, t}^2$ & 0.150** & &   0.184*** \\ 
  & (0.059) & &  (0.067) \\ [0.4cm]
  $\Delta TAX_{h ,s, t-1}$ & 0.025* & &  0.030** \\ 
  & (0.013) & &   (0.013) \\
  $\phi_{h,s,t}$ & $-$5.601*** & &  $-$5.518*** \\ 
  & (0.589) & &  (0.567) \\ 
  $\phi_{h,s,t}^2$ & 6.209*** & &  6.135*** \\ 
  & (0.813) & &  (0.797) \\ 
  $\Delta (\text{Labor cost})_{h ,s, t}$ & $-$0.051*** & &  $-$0.050** \\ 
  & (0.018) & &  (0.022) \\ 
  $\Delta (\text{Labor cost})_{h ,s, t} \cdot \phi_{h,s,t}$ & 0.405*** & &  0.402*** \\ 
  & (0.104) & &  (0.104) \\ 
  $\Delta (\text{Labor cost})_{h ,s, t} \cdot \phi_{h,s,t}^2$ & $-$0.780*** & &  $-$0.776*** \\ 
  & (0.166) & &  (0.166) \\  
  $\Delta (\text{Productivity})_{h ,s, t}$ & 0.195*** & &  0.193*** \\ 
  & (0.061) & &  (0.061) \\ 
  $\Delta (\text{Productivity})_{h,s, t} \cdot \phi_{h,s,t}$  & $-$1.296*** & &  $-$1.283*** \\ 
  & (0.381) & &  (0.382) \\ 
  $\Delta (\text{Productivity})_{h,s, t} \cdot \phi_{h,s,t}^2$  & 2.157*** & &  2.142*** \\ 
  & (0.549) & &  (0.550) \\ 
  $\Delta (\text{Pro business})_{h ,s, t}$& 0.026* & &  0.022 \\ 
  & (0.016) & &  (0.015) \\ 
  $\Delta (\text{Pro business})_{h ,s, t} \cdot \phi_{h,s,t}$ & $-$0.305*** & &  $-$0.280*** \\ 
  & (0.099) & &  (0.091) \\ 
  $\Delta (\text{Pro business})_{h ,s, t} \cdot \phi_{h,s,t}^2$ & 0.624*** & &  0.581*** \\ 
  & (0.162) & &  (0.149) \\ \hline
	Host country--year dummy & $\checkmark$ & &  $\checkmark$ \\
	Source country--year dummy & $\checkmark$ & &  $\checkmark$ \\ 
	Year dummy & $\checkmark$ & &  $\checkmark$  \\ 
	Observations & 1,352 & &  1,352 \\ 
	$R^2$ & 0.583 & &  0.583 \\ \hline \hline 
\end{tabular} \\ 
\begin{flushleft} {\footnotesize
\textit{Notes}: The dependent variable is the share of a source country's manufacturing affiliates in a host country out of the source's affiliates in all sectors in the host in a year.
Standard errors clustered at the host country-year level are in parentheses.
Unlike Table 1, we use the inverse of the log of the Head-Ries index as a measure of trade openness $\phi_{h,s,t}$. \\
***Significant at the 1$\%$ level; **Significant at the 5$\%$ level; *Significant at the 10$\%$ level.}
\end{flushleft}
\end{center}
\endgroup

\

\begin{spacing}{1.0}
\subsection*{Appendix 4. \ Parameter values}

The figures in the text were produced using the following parameter values: 

\noindent
Figure 2: $\sigma=5$, $t_1=0.3$, $t_2=0.2$, $\delta=0$, $s_1=0.5$.

\noindent
Figure 3: $\sigma=5$, $t_1=0.3$, $t_2=0.2$, $\delta \in \{ 0, 0.07, 0.143\}$, $s_1=0.5$.

\noindent
Figure 4: $\sigma=5$, $\alpha_1=4$, $\alpha_2=0.33$, $\delta=0$, $L=20$, $s_1=0.5$, $\mu=1$.

The figures do not depend on other parameter values listed above.
The choice of elasticity of substitution ($\sigma=5$) is the lower bound of the range estimated by \cite{LaiTrefler2002} and is also close to $4$, the median estimate by \cite{BrodaWeinstein2006}.
The choice of tax rates ($t_1=0.3$; $t_2=0.2$) is in line with the average tax rate is $0.2747$ for our sample of 23 OECD countries in 2008 to 2016.

\end{spacing}

\

\

\begin{spacing}{1.0}
\bibliographystyle{apalike}
\bibliography{C:/Users/Hayato/Dropbox/TeX/texlive/2019/texmf-dist/bibtex/econ_ref_hayato.bib}

\begin{thebibliography}{}

\bibitem[Agrawal and Wildasin, 2020]{AgrawalWildasin2020}
Agrawal, D.~R. and Wildasin, D.~E. (2020).
\newblock Technology and tax systems.
\newblock {\em Journal of Public Economics}, 185:104082.

\bibitem[Amerighi and Peralta, 2010]{AmerighiPeralta2010}
Amerighi, O. and Peralta, S. (2010).
\newblock The proximity-concentration trade-off with profit shifting.
\newblock {\em Journal of Urban Economics}, 68(1):90--101.

\bibitem[Andersson and Forslid, 2003]{AnderssonForslid2003}
Andersson, F. and Forslid, R. (2003).
\newblock Tax competition and economic geography.
\newblock {\em Journal of Public Economic Theory}, 5(2):279--303.

\bibitem[Antr{\`a}s and Helpman, 2004]{AntrasHelpman2004}
Antr{\`a}s, P. and Helpman, E. (2004).
\newblock Global sourcing.
\newblock {\em Journal of Political Economy}, 112(3):552--580.

\bibitem[Arvis et~al., 2016]{Arvisetal2016}
Arvis, J.-F., Duval, Y., Shepherd, B., Utoktham, C., and Raj, A. (2016).
\newblock Trade costs in the developing world: 1996--2010.
\newblock {\em World Trade Review}, 15(3):451--474.

\bibitem[Baldwin and Krugman, 2004]{BaldwinKrugman2004}
Baldwin, R.~E. and Krugman, P.~R. (2004).
\newblock Agglomeration, integration and tax harmonisation.
\newblock {\em European Economic Review}, 48(1):1--23.

\bibitem[Baldwin and Okubo, 2014]{BaldwinOkubo2014SEA}
Baldwin, R.~E. and Okubo, T. (2014).
\newblock Tax competition with heterogeneous firms.
\newblock {\em Spatial Economic Analysis}, 9(3):309--326.

\bibitem[Bartelsman and Beetsma, 2003]{BartelsmanBeestma2003}
Bartelsman, E.~J. and Beetsma, R.~M. (2003).
\newblock Why pay more? {C}orporate tax avoidance through transfer pricing in
  \text{OECD} countries.
\newblock {\em Journal of Public Economics}, 87(9):2225--2252.

\bibitem[Bauer and Langenmayr, 2013]{BauerLangenmayr2013}
Bauer, C.~J. and Langenmayr, D. (2013).
\newblock Sorting into outsourcing: Are profits taxed at a gorilla's arm's
  length?
\newblock {\em Journal of International Economics}, 90(2):326--336.

\bibitem[Becker and Riedel, 2013]{BeckerRiedel2013}
Becker, J. and Riedel, N. (2013).
\newblock Multinational firms mitigate tax competition.
\newblock {\em Economics Letters}, 118(2):404--406.

\bibitem[Beer et~al., 2020]{Beeretal2020}
Beer, S., De~Mooij, R., and Liu, L. (2020).
\newblock International corporate tax avoidance: A review of the channels,
  magnitudes, and blind spots.
\newblock {\em Journal of Economic Surveys}, 34(3):660--688.

\bibitem[Behrens et~al., 2014]{Behrensetal2014}
Behrens, K., Peralta, S., and Picard, P.~M. (2014).
\newblock Transfer pricing rules, {OECD} guidelines, and market distortions.
\newblock {\em Journal of Public Economic Theory}, 16(4):650--680.

\bibitem[B{\'e}nassy-Qu{\'e}r{\'e} et~al., 2005]{Benassyetal2005}
B{\'e}nassy-Qu{\'e}r{\'e}, A., Fontagn{\'e}, L., and Lahr{\`e}che-R{\'e}vil, A.
  (2005).
\newblock How does {FDI} react to corporate taxation?
\newblock {\em International Tax and Public Finance}, 12(5):583--603.

\bibitem[Bernard et~al., 2006]{Bernardetal2006}
Bernard, A.~B., Jensen, J.~B., and Schott, P.~K. (2006).
\newblock Transfer pricing by \text{US}-based multinational firms.
\newblock NBER Working Paper, 12493.

\bibitem[Bilir et~al., 2019]{Biliretal2019}
Bilir, L.~K., Chor, D., and Manova, K. (2019).
\newblock Host-country financial development and multinational activity.
\newblock {\em European Economic Review}, 115:192--220.

\bibitem[Blonigen and Piger, 2014]{BlonigenPiger2014}
Blonigen, B.~A. and Piger, J. (2014).
\newblock Determinants of foreign direct investment.
\newblock {\em Canadian Journal of Economics}, 47(3):775--812.

\bibitem[Blouin and Robinson, 2020]{BlouinRobinson2020}
Blouin, J.~L. and Robinson, L.~A. (2020).
\newblock Double counting accounting: How much profit of multinational
  enterprises is really in tax havens?
\newblock mimeo.

\bibitem[Bond and Gresik, 2020]{BondGresik2020}
Bond, E.~W. and Gresik, T.~A. (2020).
\newblock Unilateral tax reform: Border adjusted taxes, cash flow taxes, and
  transfer pricing.
\newblock {\em Journal of Public Economics}, 184:104160.

\bibitem[Borck and Pfl{\"u}ger, 2006]{BorckPfluger2006}
Borck, R. and Pfl{\"u}ger, M. (2006).
\newblock Agglomeration and tax competition.
\newblock {\em European Economic Review}, 50(3):647--668.

\bibitem[Broda and Weinstein, 2006]{BrodaWeinstein2006}
Broda, C. and Weinstein, D.~E. (2006).
\newblock Globalization and the gains from variety.
\newblock {\em Quarterly Journal of Economics}, 121(2):541--585.

\bibitem[Bruner et~al., 2018]{Bruneretal2018}
Bruner, J., Rassier, D.~G., and Ruhl, K.~J. (2018).
\newblock Multinational profit shifting and measures throughout economic
  accounts.
\newblock In {\em The Challenges of Globalization in the Measurement of
  National Accounts}. University of Chicago Press.

\bibitem[Bucovetsky, 1991]{Bucovetsky1991}
Bucovetsky, S. (1991).
\newblock Asymmetric tax competition.
\newblock {\em Journal of Urban Economics}, 30(2):167--181.

\bibitem[Choe and Matsushima, 2013]{ChoeMatsushima2013}
Choe, C. and Matsushima, N. (2013).
\newblock The arm's length principle and tacit collusion.
\newblock {\em International Journal of Industrial Organization},
  31(1):119--130.

\bibitem[Choi et~al., 2020]{Choietal2020}
Choi, J.~P., Furusawa, T., and Ishikawa, J. (2020).
\newblock Transfer pricing regulation and tax competition.
\newblock {\em Journal of International Economics}, 127:103367.

\bibitem[Choi et~al., 2019]{Choietal2019}
Choi, J.~P., Ishikawa, J., and Okoshi, H. (2019).
\newblock Tax havens and cross-border licensing.
\newblock RIETI Discussion Paper Series 19-E-105.

\bibitem[Choi et~al., 2021]{Choietal2021}
Choi, S., Furceri, D., and Yoon, C. (2021).
\newblock Policy uncertainty and foreign direct investment.
\newblock {\em Review of International Economics}, 29(2):195--227.

\bibitem[Clausing, 2003]{Clausing2003}
Clausing, K.~A. (2003).
\newblock Tax-motivated transfer pricing and \text{US} intrafirm trade prices.
\newblock {\em Journal of Public Economics}, 87(9):2207--2223.

\bibitem[Copithorne, 1971]{Copithorne1971}
Copithorne, L.~W. (1971).
\newblock International corporate transfer prices and government policy.
\newblock {\em Canadian Journal of Economics}, 4(3):324--341.

\bibitem[Cristea and Nguyen, 2016]{CristeaNguyen2016}
Cristea, A.~D. and Nguyen, D.~X. (2016).
\newblock Transfer pricing by multinational firms: New evidence from foreign
  firm ownerships.
\newblock {\em American Economic Journal: Economic Policy}, 8(3):170--202.

\bibitem[Da~Rin et~al., 2010]{DaRinetal2010}
Da~Rin, M., Di~Giacomo, M., and Sembenelli, A. (2010).
\newblock Corporate taxation and the size of new firms: Evidence from {E}urope.
\newblock {\em Journal of the European Economic Association}, 8(2-3):606--616.

\bibitem[Da~Rin et~al., 2011]{DaRinetal2011}
Da~Rin, M., Di~Giacomo, M., and Sembenelli, A. (2011).
\newblock Entrepreneurship, firm entry, and the taxation of corporate income:
  Evidence from {E}urope.
\newblock {\em Journal of Public Economics}, 95(9-10):1048--1066.

\bibitem[Darby et~al., 2014]{Darbyetal2014}
Darby, J., Ferrett, B., and Wooton, I. (2014).
\newblock Regional centrality and tax competition for \text{FDI}.
\newblock {\em Regional Science and Urban Economics}, 49:84--92.

\bibitem[Davies and Eckel, 2010]{DaviesEckel2010}
Davies, R.~B. and Eckel, C. (2010).
\newblock Tax competition for heterogeneous firms with endogenous entry.
\newblock {\em American Economic Journal: Economic Policy}, 2(1):77--102.

\bibitem[Davies et~al., 2018]{Daviesetal2018}
Davies, R.~B., Martin, J., Parenti, M., and Toubal, F. (2018).
\newblock Knocking on tax haven's door: Multinational firms and transfer
  pricing.
\newblock {\em Review of Economics and Statistics}, 100(1):120--134.

\bibitem[Devereux and Griffith, 1999]{DevereuxGriffith1999}
Devereux, M.~P. and Griffith, R. (1999).
\newblock The taxation of discrete investment choices.
\newblock Working Paper Series No. W98/16, The Institute for Fiscal Studies.

\bibitem[Dharmapala, 2014]{Dharmapala2014}
Dharmapala, D. (2014).
\newblock What do we know about base erosion and profit shifting? {A} review of
  the empirical literature.
\newblock {\em Fiscal Studies}, 35(4):421--448.

\bibitem[Dixit and Stiglitz, 1977]{DixitStiglitz1977}
Dixit, A. and Stiglitz, J. (1977).
\newblock Monopolistic competition and optimum product diversity.
\newblock {\em American Economic Review}, 67(3):297--308.

\bibitem[Egger et~al., 2009]{Eggeretal2009}
Egger, P., Loretz, S., Pfaffermayr, M., and Winner, H. (2009).
\newblock Bilateral effective tax rates and foreign direct investment.
\newblock {\em International Tax and Public Finance}, 16(6):822--849.

\bibitem[Egger and Seidel, 2013]{EggerSeidel2013}
Egger, P. and Seidel, T. (2013).
\newblock Corporate taxes and intra-firm trade.
\newblock {\em European Economic Review}, 63:225--242.

\bibitem[Elitzur and Mintz, 1996]{ElitzurMintz1996}
Elitzur, R. and Mintz, J. (1996).
\newblock Transfer pricing rules and corporate tax competition.
\newblock {\em Journal of Public Economics}, 60(3):401--422.

\bibitem[Fuest et~al., 2005]{Fuestetal2005}
Fuest, C., Huber, B., and Mintz, J. (2005).
\newblock Capital mobility and tax competition.
\newblock {\em Foundations and Trends in Microeconomics}, 1(1):1--62.

\bibitem[Fujita et~al., 1999]{Fujitaetal1999}
Fujita, M., Krugman, P.~R., and Venables, A.~J. (1999).
\newblock {\em The Spatial Economy}.
\newblock MIT Press, Cambridge, MA.

\bibitem[Goldbach et~al., 2019]{Goldbachetal2019}
Goldbach, S., Nagengast, A.~J., Steinm{\"u}ller, E., and Wamser, G. (2019).
\newblock The effect of investing abroad on investment at home: On the role of
  technology, tax savings, and internal capital markets.
\newblock {\em Journal of International Economics}, 116:58--73.

\bibitem[Gumpert et~al., 2016]{Gumpertetal2016}
Gumpert, A., Hines~Jr, J.~R., and Schnitzer, M. (2016).
\newblock Multinational firms and tax havens.
\newblock {\em Review of Economics and Statistics}, 98(4):713--727.

\bibitem[Guvenen et~al., 2017]{Guvenenetal2017}
Guvenen, F., Mataloni~Jr, R.~J., Rassier, D.~G., and Ruhl, K.~J. (2017).
\newblock Offshore profit shifting and domestic productivity measurement.
\newblock NBER Working Paper, 23324.

\bibitem[Habu, 2017]{Habu2017}
Habu, K. (2017).
\newblock Centre for business taxation tax database 2017.
\newblock University of Oxford.

\bibitem[Han et~al., 2014]{Hanetal2014}
Han, Y., Pieretti, P., Zanaj, S., and Zou, B. (2014).
\newblock Asymmetric competition among nation states: A differential game
  approach.
\newblock {\em Journal of Public Economics}, 119:71--79.

\bibitem[Han et~al., 2018]{Hanetal2018RIE}
Han, Y., Pieretti, P., and Zou, B. (2018).
\newblock Does tax competition increase infrastructural disparity among
  jurisdictions?
\newblock {\em Review of International Economics}, 26(1):20--36.

\bibitem[Hauck, 2019]{Hauck2019}
Hauck, T. (2019).
\newblock Lobbying and the international fight against tax havens.
\newblock {\em Journal of Public Economic Theory}, 21(3):537--557.

\bibitem[Haufler and Runkel, 2012]{HauflerRunkel2012}
Haufler, A. and Runkel, M. (2012).
\newblock Firms' financial choices and thin capitalization rules under
  corporate tax competition.
\newblock {\em European Economic Review}, 56(6):1087--1103.

\bibitem[Haufler and Schjelderup, 2000]{HauflerSchjelderup2000}
Haufler, A. and Schjelderup, G. (2000).
\newblock Corporate tax systems and cross country profit shifting.
\newblock {\em Oxford Economic Papers}, 52(2):306--325.

\bibitem[Haufler and St{\"a}hler, 2013]{HauflerStahler2013}
Haufler, A. and St{\"a}hler, F. (2013).
\newblock Tax competition in a simple model with heterogeneous firms: {H}ow
  larger markets reduce profit taxes.
\newblock {\em International Economic Review}, 54(2):665--692.

\bibitem[Head and Ries, 2001]{HeadRies2001}
Head, K. and Ries, J. (2001).
\newblock Increasing returns versus national product differentiation as an
  explanation for the pattern of \text{US-Canada} trade.
\newblock {\em American Economic Review}, 91(4):858--876.

\bibitem[Heckemeyer and Overesch, 2017]{HeckemeyerOveresch2017}
Heckemeyer, J.~H. and Overesch, M. (2017).
\newblock Multinationals' profit response to tax differentials: Effect size and
  shifting channels.
\newblock {\em Canadian Journal of Economics}, 50(4):965--994.

\bibitem[Helpman and Krugman, 1985]{HelpmanKrugman1985}
Helpman, E. and Krugman, P.~R. (1985).
\newblock {\em Market Structure and Foreign Trade}.
\newblock MIT Press, Cambridge, MA.

\bibitem[Helpman et~al., 2004]{Helpmanetal2004}
Helpman, E., Melitz, M.~J., and Yeaple, S.~R. (2004).
\newblock Export versus {FDI} with heterogeneous firms.
\newblock {\em American economic review}, 94(1):300--316.

\bibitem[Hindriks and Nishimura, 2021]{HindriksNishimura2021}
Hindriks, J. and Nishimura, Y. (2021).
\newblock Taxing multinationals: The scope for enforcement cooperation.
\newblock {\em Journal of Public Economic Theory}, 23(3):487--509.

\bibitem[Horner and Aoyama, 2009]{HornerAoyama2009}
Horner, R. and Aoyama, Y. (2009).
\newblock Limits to {FDI}-driven growth in {I}reland: A newspaper content
  analysis for investment, upgrading and divestment.
\newblock {\em Irish Geography}, 42(2):185--205.

\bibitem[Horst, 1971]{Horst1971}
Horst, T. (1971).
\newblock The theory of the multinational firm: Optimal behavior under
  different tariff and tax rates.
\newblock {\em Journal of Political Economy}, 79(5):1059--1072.

\bibitem[Janeba and Osterloh, 2013]{JanebaOsterloh2013}
Janeba, E. and Osterloh, S. (2013).
\newblock Tax and the city—{A} theory of local tax competition.
\newblock {\em Journal of Public Economics}, 106:89--100.

\bibitem[Jansk{\`y} and Palansk{\`y}, 2019]{JanskyPalansky2019}
Jansk{\`y}, P. and Palansk{\`y}, M. (2019).
\newblock Estimating the scale of profit shifting and tax revenue losses
  related to foreign direct investment.
\newblock {\em International Tax and Public Finance}, 26(5):1048--1103.

\bibitem[Johannesen, 2010]{Johannesen2010}
Johannesen, N. (2010).
\newblock Imperfect tax competition for profits, asymmetric equilibrium and
  beneficial tax havens.
\newblock {\em Journal of International Economics}, 81(2):253--264.

\bibitem[Juranek et~al., 2018]{Juraneketal2018}
Juranek, S., Schindler, D., and Schjelderup, G. (2018).
\newblock Transfer pricing regulation and taxation of royalty payments.
\newblock {\em Journal of Public Economic Theory}, 20(1):67--84.

\bibitem[Kato, 2015]{Kato2015}
Kato, H. (2015).
\newblock The importance of government commitment in attracting firms: {A}
  dynamic analysis of tax competition in an agglomeration economy.
\newblock {\em European Economic Review}, 74(C):57--78.

\bibitem[Kato and Okoshi, 2019]{KatoOkoshi2019}
Kato, H. and Okoshi, H. (2019).
\newblock Production location of multinational firms under transfer pricing:
  The impact of the arm's length principle.
\newblock {\em International Tax and Public Finance}, 26(4):835--871.

\bibitem[Keen and Konrad, 2013]{KeenKonrad2013}
Keen, M. and Konrad, K.~A. (2013).
\newblock The theory of international tax competition and coordination.
\newblock In Auerbach, A.~J., Chetty, R., Feldstein, M., and Saez, E., editors,
  {\em Handbook of Public Economics}, volume~5, pages 257--328. Elsevier,
  Oxford.

\bibitem[Keuschnigg and Devereux, 2013]{KeuschniggDevereux2013}
Keuschnigg, C. and Devereux, M.~P. (2013).
\newblock The arm's length principle and distortions to multinational firm
  organization.
\newblock {\em Journal of International Economics}, 89(2):432--440.

\bibitem[Kind et~al., 2000]{Kindetal2000}
Kind, H.~J., Knarvik, K. H.~M., and Schjelderup, G. (2000).
\newblock Competing for capital in a `lumpy' world.
\newblock {\em Journal of Public Economics}, 78(3):253--274.

\bibitem[Kind et~al., 2005]{Kindetal2005}
Kind, H.~J., Midelfart, K.~H., and Schjelderup, G. (2005).
\newblock Corporate tax systems, multinational enterprises, and economic
  integration.
\newblock {\em Journal of International Economics}, 65(2):507--521.

\bibitem[Krautheim and Schmidt-Eisenlohr, 2011]{Krautheimetal2011}
Krautheim, S. and Schmidt-Eisenlohr, T. (2011).
\newblock Heterogeneous firms, profit shifting of \text{FDI} and international
  tax competition.
\newblock {\em Journal of Public Economics}, 95(1):122--133.

\bibitem[Lai and Trefler, 2002]{LaiTrefler2002}
Lai, H. and Trefler, D. (2002).
\newblock The gains from trade with monopolistic competition: Specification,
  estimation, and mis-specification.
\newblock NBER Working Paper, 24884.

\bibitem[Langenmayr et~al., 2015]{Langenmayretal2015}
Langenmayr, D., Haufler, A., and Bauer, C.~J. (2015).
\newblock Should tax policy favor high-or low-productivity firms?
\newblock {\em European Economic Review}, 73:18--34.

\bibitem[Ludema and Wooton, 2000]{LudemaWooton2000}
Ludema, R.~D. and Wooton, I. (2000).
\newblock Economic geography and the fiscal effects of regional integration.
\newblock {\em Journal of International Economics}, 52(2):331--357.

\bibitem[Ma and Raimondos, 2015]{MaRaimondos2015}
Ma, J. and Raimondos, P. (2015).
\newblock Competition for \text{FDI} and profit shifting.
\newblock CESifo Working Paper Series, 5153.

\bibitem[Martin and Rogers, 1995]{MartinRogers1995}
Martin, P. and Rogers, C.~A. (1995).
\newblock Industrial location and public infrastructure.
\newblock {\em Journal of International Economics}, 39(3):335--351.

\bibitem[Matsui, 2012]{Matsui2012}
Matsui, K. (2012).
\newblock Auditing internal transfer prices in multinationals under
  monopolistic competition.
\newblock {\em International Tax and Public Finance}, 19(6):800--818.

\bibitem[Mayer et~al., 2010]{Mayeretal2010}
Mayer, T., M{\'e}jean, I., and Nefussi, B. (2010).
\newblock The location of domestic and foreign production affiliates by
  {F}rench multinational firms.
\newblock {\em Journal of Urban Economics}, 68(2):115--128.

\bibitem[Mayer and Zignago, 2011]{MayerZignago2011}
Mayer, T. and Zignago, S. (2011).
\newblock Notes on {CEPII}'s distances measures: The {GeoDist} database.
\newblock CEPII Working Paper 2011-25.

\bibitem[Navaretti and Venables, 2004]{NavarettiVenables2004}
Navaretti, G.~B. and Venables, A. (2004).
\newblock {\em Multinational Firms in the World Economy}.
\newblock Princeton University Press, Princeton, NJ.

\bibitem[Nielsen et~al., 2003]{Nielsenetal2003}
Nielsen, S.~B., Raimondos-M{\o}ller, P., and Schjelderup, G. (2003).
\newblock Formula apportionment and transfer pricing under oligopolistic
  competition.
\newblock {\em Journal of Public Economic Theory}, 5(2):419--437.

\bibitem[Nielsen et~al., 2008]{Nielsenetal2008}
Nielsen, S.~B., Raimondos-M{\o}ller, P., and Schjelderup, G. (2008).
\newblock Taxes and decision rights in multinationals.
\newblock {\em Journal of Public Economic Theory}, 10(2):245--258.

\bibitem[Novy, 2013]{Novy2013}
Novy, D. (2013).
\newblock Gravity redux: Measuring international trade costs with panel data.
\newblock {\em Economic Inquiry}, 51(1):101--121.

\bibitem[OECD, 2017]{OECD2017admin}
OECD (2017).
\newblock {\em Tax Administration 2017: Comparative Information on {OECD} and
  Other Advanced and Emerging Economies}.
\newblock OECD, Paris.

\bibitem[Okoshi, 2020]{OkoshiPHD}
Okoshi, H. (2020).
\newblock {\em Transfer Pricing: Interaction with Multionationals' Location,
  Export and R\&D Choices}.
\newblock Ph.D. dissertation submitted to the University of Munich.

\bibitem[Ottaviano and van Ypersele, 2005]{OttavianoYpersele2005}
Ottaviano, G.~I. and van Ypersele, T. (2005).
\newblock Market size and tax competition.
\newblock {\em Journal of International Economics}, 67(1):25--46.

\bibitem[Overesch, 2009]{Overesch2009}
Overesch, M. (2009).
\newblock The effects of multinationals' profit shifting activities on real
  investments.
\newblock {\em National Tax Journal}, 62(1):5--23.

\bibitem[Peralta et~al., 2006]{Peraltaetal2006}
Peralta, S., Wauthy, X., and van Ypersele, T. (2006).
\newblock Should countries control international profit shifting?
\newblock {\em Journal of International Economics}, 68(1):24--37.

\bibitem[Pfl{\"u}ger, 2004]{Pfluger2004}
Pfl{\"u}ger, M. (2004).
\newblock A simple, analytically solvable, {C}hamberlinian agglomeration model.
\newblock {\em Regional Science and Urban Economics}, 34(5):565--573.

\bibitem[Profeta and Scabrosetti, 2017]{ProfetaScabrosetti2017}
Profeta, P. and Scabrosetti, S. (2017).
\newblock The political economy of taxation in {Europe}.
\newblock {\em Hacienda P{\'u}blica Espa{\~n}ola}, 1(220):139--172.

\bibitem[Schjelderup and S{\o}rgard, 1997]{SchjelderupSorgard1997}
Schjelderup, G. and S{\o}rgard, L. (1997).
\newblock Transfer pricing as a strategic device for decentralized
  multinationals.
\newblock {\em International Tax and Public Finance}, 4(3):277--290.

\bibitem[Shen, 2018]{Shen2018}
Shen, Y. (2018).
\newblock Corporate income taxation and multinational production.
\newblock mimeo. Hanyang University.

\bibitem[Slemrod and Wilson, 2009]{SlemrodWilson2009}
Slemrod, J. and Wilson, J.~D. (2009).
\newblock Tax competition with parasitic tax havens.
\newblock {\em Journal of Public Economics}, 93(11-12):1261--1270.

\bibitem[St{\"o}whase, 2005]{Stowhase2005}
St{\"o}whase, S. (2005).
\newblock Asymmetric capital tax competition with profit shifting.
\newblock {\em Journal of Economics}, 85(2):175--196.

\bibitem[St{\"o}whase, 2013]{Stowhase2013}
St{\"o}whase, S. (2013).
\newblock How profit shifting may increase the tax burden of multinationals: A
  simple model with discrete investment choices.
\newblock {\em Journal of Public Economic Theory}, 15(2):185--207.

\bibitem[Swenson, 2001]{Swenson2001}
Swenson, D.~L. (2001).
\newblock Tax reforms and evidence of transfer pricing.
\newblock {\em National Tax Journal}, 54(1):7--25.

\bibitem[T{\o}rsl{\o}v et~al., 2018]{Torslovetal2020}
T{\o}rsl{\o}v, T.~R., Wier, L.~S., and Zucman, G. (2018).
\newblock The missing profits of nations.
\newblock mimeo.

\bibitem[Wang, 2020]{Wang2020}
Wang, Z. (2020).
\newblock Multinational production and corporate taxes: A quantitative
  assessment.
\newblock {\em Journal of International Economics}, 126:103353.

\bibitem[Wilson, 1991]{Wilson1991}
Wilson, J.~D. (1991).
\newblock Tax competition with interregional differences in factor endowments.
\newblock {\em Regional Science and Urban Economics}, 21(3):423--451.

\bibitem[Yao, 2013]{Yao2013}
Yao, J.~T. (2013).
\newblock The arm's length principle, transfer pricing, and location choices.
\newblock {\em Journal of Economics and Business}, 65:1--13.

\bibitem[Zhao, 2000]{Zhao2000}
Zhao, L. (2000).
\newblock Decentralization and transfer pricing under oligopoly.
\newblock {\em Southern Economic Journal}, 67(2):414--426.

\bibitem[Ziss, 2007]{Ziss2007}
Ziss, S. (2007).
\newblock Hierarchies, intra-firm competition and mergers.
\newblock {\em International Journal of Industrial Organization},
  25(2):237--260.

\bibitem[Zucman, 2014]{Zucman2014}
Zucman, G. (2014).
\newblock Taxing across borders: {T}racking personal wealth and corporate
  profits.
\newblock {\em Journal of Economic Perspectives}, 28(4):121--48.

\end{thebibliography}
\end{spacing}

\end{document}